
%
%
%
%
\message{The figures are contained in a PostScript file which must be
         printed separately.}
%
%

\font\eightrm=cmr8 at 8pt

\font\seventeenrm=cmr17 at 17pt
\font\twentyonerm=cmr17 at 21pt

\font\ss=cmss10

\font\csc=cmcsc10

\font\twelvecal=cmsy10 at 12pt

\font\twelvemath=cmmi12

\font\seventeenbold=cmbx7 at 17pt

\font\fively=lasy5
\font\sevenly=lasy7
\font\tenly=lasy10

\textfont10=\tenly
\scriptfont10=\sevenly
\scriptscriptfont10=\fively
\magnification=1200
\parskip=10pt
\parindent=20pt
\def\today{\ifcase\month\or January\or February\or March\or April\or May\or
June
       \or July\or August\or September\or October\or November\or December\fi
       \space\number\day, \number\year}

\def\title#1{\footline={\ifnum\pageno<2\hfil
       \else\hss\tenrm\folio\hss\fi}\vskip1truein\centerline{{#1}
       \footnote{\raise1ex\hbox{*}}{\eightrm Supported in part
       by the Robert A. Welch Foundation and N.S.F. Grants
       PHY-880637 and\break PHY-8605978.}}}

\def\newpage{\vfill\eject}
\def\abstract#1{\centerline{\bf ABSTRACT}\vskip.2truein{\narrower\noindent#1
       \smallskip}}
\def\acknowledgements{\noindent\line{\bf Acknowledgements\hfill}\nobreak
    \vskip.1truein\nobreak\noindent\ignorespaces}
\def\runninghead#1#2{\voffset=2\baselineskip\nopagenumbers
       \headline={\ifodd\pageno\rightheadline\else \leftheadline\fi}
       \def\rightheadline{{\sl#1}\hfill{\rm\folio}}
       \def\leftheadline{{\rm\folio}\hfill{\sl#2}}}
\def\SS{\mathhexbox278}

\newcount\footnoteno
\def\Footnote#1{\advance\footnoteno by 1
                \let\SF=\empty
                \ifhmode\edef\SF{\spacefactor=\the\spacefactor}\/\fi
                $^{\the\footnoteno}$\ignorespaces
                \SF\vfootnote{$^{\the\footnoteno}$}{#1}}

\def\place#1#2#3{\vbox to0pt{\kern-\parskip\kern-7pt
                             \kern-#2truein\hbox{\kern#1truein #3}
                             \vss}\nointerlineskip}
\def\figurecaption#1#2{\kern.75truein\vbox{\hsize=5truein\noindent{\bf Figure
    \figlabel{#1}:} #2}}
\def\tablecaption#1#2{\kern.75truein\lower12truept\hbox{\vbox{\hsize=5truein
    \noindent{\bf Table\hskip5truept\tablabel{#1}:} #2}}}
\def\boxed#1{\lower3pt\hbox{
                       \vbox{\hrule\hbox{\vrule
                        \vbox{\kern2pt\hbox{\kern3pt#1\kern3pt}\kern3pt}\vrule}
                         \hrule}}}
\def\a{\alpha}
\def\b{\beta}
\def\g{\gamma}\def\G{\Gamma}
\def\d{\delta}\def\D{\Delta}
\def\e{\epsilon}
\def\z{\zeta}

\def\th{\theta}

\def\l{\lambda}

\def\n{\nu}
\def\x{\xi}

\def\p{\pi}\def\P{\Pi}\def\vp{\varpi}

\def\s{\sigma}\def\S{\Sigma}
\def\t{\tau}

\def\ph{\phi}
\def\ch{\chi}
\def\ps{\psi}\def\Ps{\Psi}
\def\o{\omega}\def\O{\Omega}

\def\ca#1{\relax\ifmmode {{\cal #1}}\else $\cal #1$\fi}

\def\calb{{\cal B}}

\def\calm{{\cal M}}

\def\inbar{\vrule height1.5ex width.4pt depth0pt}
\def\IB{\relax{\rm I\kern-.18em B}}
\def\IC{\relax\hbox{\kern.25em$\inbar\kern-.3em{\rm C}$}}
\def\ID{\relax{\rm I\kern-.18em D}}
\def\IE{\relax{\rm I\kern-.18em E}}
\def\IF{\relax{\rm I\kern-.18em F}}
\def\IG{\relax\hbox{\kern.25em$\inbar\kern-.3em{\rm G}$}}
\def\IH{\relax{\rm I\kern-.18em H}}
\def\II{\relax{\rm I\kern-.18em I}}
\def\IK{\relax{\rm I\kern-.18em K}}
\def\IL{\relax{\rm I\kern-.18em L}}
\def\IM{\relax{\rm I\kern-.18em M}}
\def\IN{\relax{\rm I\kern-.18em N}}
\def\IO{\relax\hbox{\kern.25em$\inbar\kern-.3em{\rm O}$}}
\def\IP{\relax{\rm I\kern-.18em P}}
\def\IQ{\relax\hbox{\kern.25em$\inbar\kern-.3em{\rm Q}$}}
\def\IR{\relax{\rm I\kern-.18em R}}
\def\IZ{\relax\ifmmode\hbox{\ss Z\kern-.4em Z}\else{\ss Z\kern-.4em Z}\fi}
\def\IGa{\relax{\rm I}\kern-.18em\Gamma}
\def\IPi{\relax{\rm I}\kern-.18em\Pi}
\def\ITh{\relax\hbox{\kern.25em$\inbar\kern-.3em\Theta$}}
\def\IOm{\relax\thinspace\inbar\kern1.95pt\inbar\kern-5.525pt\Omega}


\def\ie{{\it i.e.\ \/}}

\def\noblackboxes{\overfullrule=0pt}
\def\define{\buildrel\rm def\over =}

\def\cy{Calabi--Yau}
\def\cym{Calabi--Yau manifold}
\def\cys{Calabi--Yau manifolds}

\def\K{K\"ahler}

\def\H#1#2{\relax\ifmmode {H^{#1#2}}\else $H^{#1 #2}$\fi}
\def\M{\relax\ifmmode{\calm}\else $\calm$\fi}

\def\Bigcheck{\lower3.8pt\hbox{\smash{\hbox{{\twentyonerm \v{}}}}}}
\def\bigboldcheck{\smash{\hbox{{\seventeenbold\v{}}}}}

\def\Bighat{\lower3.8pt\hbox{\smash{\hbox{{\twentyonerm \^{}}}}}}

\def\Msharp{\relax\ifmmode{\calm^\sharp}\else $\smash{\calm^\sharp}$\fi}
\def\Mflat{\relax\ifmmode{\calm^\flat}\else $\smash{\calm^\flat}$\fi}
\def\preMcheck{\kern2pt\hbox{\Bigcheck\kern-12pt{$\cal M$}}}
\def\Mcheck{\relax\ifmmode\preMcheck\else $\preMcheck$\fi}
\def\preMhat{\kern2pt\hbox{\Bighat\kern-12pt{$\cal M$}}}
\def\Mhat{\relax\ifmmode\preMhat\else $\preMhat$\fi}

\def\Bsharp{\relax\ifmmode{\calb^\sharp}\else $\calb^\sharp$\fi}
\def\Bflat{\relax\ifmmode{\calb^\flat}\else $\calb^\flat$ \fi}
\def\preBcheck{\hbox{\Bigcheck\kern-9pt{$\cal B$}}}
\def\Bcheck{\relax\ifmmode\preBcheck\else $\preBcheck$\fi}
\def\preBhat{\hbox{\Bighat\kern-9pt{$\cal B$}}}
\def\Bhat{\relax\ifmmode\preBhat\else $\preBhat$\fi}

\def\figBcheck{\kern3pt\hbox{\raise1pt\hbox{\bigboldcheck}\kern-11pt
    {\twelvecal B}}}
\def\figBsharp{{\twelvecal B}\raise5pt\hbox{$\twelvemath\sharp$}}
\def\figBflat{{\twelvecal B}\raise5pt\hbox{$\twelvemath\flat$}}

\def\gcheck{\hbox{\lower2.5pt\hbox{\Bigcheck}\kern-8pt$\g$}}
\def\lhat{\hbox{\raise.5pt\hbox{\Bighat}\kern-8pt$\l$}}

\def\Fcheck{\kern2pt\hbox{\raise1pt\hbox{\Bigcheck}\kern-10pt{$\cal F$}}}
\def\Fhat{\kern2pt\hbox{\raise1pt\hbox{\Bighat}\kern-10pt{$\cal F$}}}

\def\cp#1{\relax\ifmmode {\IP\kern-2pt{}_{#1}}\else $\IP\kern-2pt{}_{#1}$\fi}
\def\h#1#2{\relax\ifmmode {b_{#1#2}}\else $b_{#1#2}$\fi}

\def\imag{\Im m}
\def\half{{1\over 2}}

\def\frac#1#2{{#1\over #2}}

\def\pd#1#2{{\partial #1\over\partial #2}}

\def\cone{\relax\thinspace\hbox{$<\kern-.8em{)}$}}
\mathchardef\mho"0A30

\def\asymp{\sim}
\def\-{\hphantom{-}}

\def\ip{\amalg}


\def\npb#1{Nucl.\ Phys.\ {\bf B#1}}

\def\plb#1{Phys. Lett. {\bf #1B}}


\def\picture #1 by #2 (#3){\vbox to #2{\hrule width #1 height 0pt depth 0pt
                                       \vfill\special{picture #3}}}
\def\scaledpicture #1 by #2 (#3 scaled #4){{\dimen0=#1 \dimen1=#2
           \divide\dimen0 by 1000 \multiply\dimen0 by #4
            \divide\dimen1 by 1000 \multiply\dimen1 by #4
            \picture \dimen0 by \dimen1 (#3 scaled #4)}}
\def\illustration #1 by #2 (#3){\vbox to #2%
{\hrule width #1 height 0pt depth0pt\vfill\special{illustration #3}}}

\def\scaledillustration #1 by #2 (#3 scaled #4){{\dimen0=#1 \dimen1=#2
           \divide\dimen0 by 1000 \multiply\dimen0 by #4
            \divide\dimen1 by 1000 \multiply\dimen1 by #4
            \illustration \dimen0 by \dimen1 (#3 scaled #4)}}


\def\delaOssa{\nobreak\vskip1truein\hbox to\hsize
       {\hskip 4truein Xenia de la Ossa\hfill}}

\def\hoy{\number\day\space de \ifcase\month\or enero\or febrero\or marzo\or
       abril\or mayo\or junio\or julio\or agosto\or septiembre\or octubre\or
       noviembre\or diciembre\fi\space de \number\year}


\newif\ifproofmode
\proofmodefalse

\newif\ifforwardreference
\forwardreferencefalse

\newif\ifchapternumbers
\chapternumbersfalse

\newif\ifcontinuousnumbering
\continuousnumberingfalse

\newif\iffigurechapternumbers
\figurechapternumbersfalse

\newif\ifcontinuousfigurenumbering
\continuousfigurenumberingfalse

\newif\iftablechapternumbers
\tablechapternumbersfalse

\newif\ifcontinuoustablenumbering
\continuoustablenumberingfalse

\font\eqsixrm=cmr6

\def\marginstyle{\eqsixrm}

\newtoks\chapletter
\newcount\chapno
\newcount\eqlabelno
\newcount\figureno
\newcount\tableno

\chapno=0
\eqlabelno=0
\figureno=0
\tableno=0

\def\chapfolio{\ifnum\chapno>0 \the\chapno\else\the\chapletter\fi}

\def\bumpchapno{\ifnum\chapno>-1 \global\advance\chapno by 1
\else\global\advance\chapno by -1 \setletter\chapno\fi
\ifcontinuousnumbering\else\global\eqlabelno=0 \fi
\ifcontinuousfigurenumbering\else\global\figureno=0 \fi
\ifcontinuoustablenumbering\else\global\tableno=0 \fi}

\def\setletter#1{\ifcase-#1{}\or{}%
\or\global\chapletter={A}%
\or\global\chapletter={B}%
\or\global\chapletter={C}%
\or\global\chapletter={D}%
\or\global\chapletter={E}%
\or\global\chapletter={F}%
\or\global\chapletter={G}%
\or\global\chapletter={H}%
\or\global\chapletter={I}%
\or\global\chapletter={J}%
\or\global\chapletter={K}%
\or\global\chapletter={L}%
\or\global\chapletter={M}%
\or\global\chapletter={N}%
\or\global\chapletter={O}%
\or\global\chapletter={P}%
\or\global\chapletter={Q}%
\or\global\chapletter={R}%
\or\global\chapletter={S}%
\or\global\chapletter={T}%
\or\global\chapletter={U}%
\or\global\chapletter={V}%
\or\global\chapletter={W}%
\or\global\chapletter={X}%
\or\global\chapletter={Y}%
\or\global\chapletter={Z}\fi}

\def\tempsetletter#1{\ifcase-#1{}\or{}%
\or\global\chapletter={A}%
\or\global\chapletter={B}%
\or\global\chapletter={C}%
\or\global\chapletter={D}%
\or\global\chapletter={E}%
\or\global\chapletter={F}%
\or\global\chapletter={G}%
\or\global\chapletter={H}%
\or\global\chapletter={I}%
\or\global\chapletter={J}%
\or\global\chapletter={K}%
\or\global\chapletter={L}%
\or\global\chapletter={M}%
\or\global\chapletter={N}%
\or\global\chapletter={O}%
\or\global\chapletter={P}%
\or\global\chapletter={Q}%
\or\global\chapletter={R}%
\or\global\chapletter={S}%
\or\global\chapletter={T}%
\or\global\chapletter={U}%
\or\global\chapletter={V}%
\or\global\chapletter={W}%
\or\global\chapletter={X}%
\or\global\chapletter={Y}%
\or\global\chapletter={Z}\fi}

\def\chapshow#1{\ifnum#1>0 \relax#1%
\else{\tempsetletter{\number#1}\chapno=#1\chapfolio}\fi}

\def\chaplabel#1{\bumpchapno\ifproofmode\ifforwardreference
\immediate\write1{\noexpand\expandafter\noexpand\def
\noexpand\csname CHAPLABEL#1\endcsname{\the\chapno}}\fi\fi
\global\expandafter\edef\csname CHAPLABEL#1\endcsname
{\the\chapno}\ifproofmode\llap{\hbox{\marginstyle #1\ }}\fi\chapfolio}

\def\chapref#1{\ifundefined{CHAPLABEL#1}??\ifproofmode\ifforwardreference%
\else\write16{ ***Undefined Chapter Reference #1*** }\fi
\else\write16{ ***Undefined Chapter Reference #1*** }\fi
\else\edef\LABxx{\getlabel{CHAPLABEL#1}}\chapshow\LABxx\fi
\ifproofmode\write2{Chapter #1}\fi}

\def\eqnum{\global\advance\eqlabelno by 1
\eqno(\ifchapternumbers\chapfolio.\fi\the\eqlabelno)}

\def\eqlabel#1{\global\advance\eqlabelno by 1 \ifproofmode\ifforwardreference
\immediate\write1{\noexpand\expandafter\noexpand\def
\noexpand\csname EQLABEL#1\endcsname{\the\chapno.\the\eqlabelno?}}\fi\fi
\global\expandafter\edef\csname EQLABEL#1\endcsname
{\the\chapno.\the\eqlabelno?}\eqno(\ifchapternumbers\chapfolio.\fi
\the\eqlabelno)\ifproofmode\rlap{\hbox{\marginstyle #1}}\fi}

\def\eqalignnum{\global\advance\eqlabelno by 1
&(\ifchapternumbers\chapfolio.\fi\the\eqlabelno)}

\def\eqalignlabel#1{\global\advance\eqlabelno by 1 \ifproofmode
\ifforwardreference\immediate\write1{\noexpand\expandafter\noexpand\def
\noexpand\csname EQLABEL#1\endcsname{\the\chapno.\the\eqlabelno?}}\fi\fi
\global\expandafter\edef\csname EQLABEL#1\endcsname
{\the\chapno.\the\eqlabelno?}&(\ifchapternumbers\chapfolio.\fi
\the\eqlabelno)\ifproofmode\rlap{\hbox{\marginstyle #1}}\fi}

\def\eqref#1{\hbox{(\ifundefined{EQLABEL#1}***)\ifproofmode\ifforwardreference%
\else\write16{ ***Undefined Equation Reference #1*** }\fi
\else\write16{ ***Undefined Equation Reference #1*** }\fi
\else\edef\LABxx{\getlabel{EQLABEL#1}}%
\def\LAByy{\expandafter\stripchap\LABxx}\ifchapternumbers%
\chapshow{\LAByy}.\expandafter\stripeq\LABxx%
\else\ifnum\number\LAByy=\chapno\relax\expandafter\stripeq\LABxx%
\else\chapshow{\LAByy}.\expandafter\stripeq\LABxx\fi\fi)\fi}%
\ifproofmode\write2{Equation #1}\fi}

\def\fignum{\global\advance\figureno by 1
\relax\iffigurechapternumbers\chapfolio.\fi\the\figureno}

\def\figlabel#1{\global\advance\figureno by 1
\relax\ifproofmode\ifforwardreference
\immediate\write1{\noexpand\expandafter\noexpand\def
\noexpand\csname FIGLABEL#1\endcsname{\the\chapno.\the\figureno?}}\fi\fi
\global\expandafter\edef\csname FIGLABEL#1\endcsname
{\the\chapno.\the\figureno?}\iffigurechapternumbers\chapfolio.\fi
\ifproofmode\llap{\hbox{\marginstyle#1
\kern1.2truein}}\relax\fi\the\figureno}

\def\figref#1{\hbox{\ifundefined{FIGLABEL#1}!!!\ifproofmode\ifforwardreference%
\else\write16{ ***Undefined Figure Reference #1*** }\fi
\else\write16{ ***Undefined Figure Reference #1*** }\fi
\else\edef\LABxx{\getlabel{FIGLABEL#1}}%
\def\LAByy{\expandafter\stripchap\LABxx}\iffigurechapternumbers%
\chapshow{\LAByy}.\expandafter\stripeq\LABxx%
\else\ifnum \number\LAByy=\chapno\relax\expandafter\stripeq\LABxx%
\else\chapshow{\LAByy}.\expandafter\stripeq\LABxx\fi\fi\fi}%
\ifproofmode\write2{Figure #1}\fi}

\def\tabnum{\global\advance\tableno by 1
\relax\iftablechapternumbers\chapfolio.\fi\the\tableno}

\def\tablabel#1{\global\advance\tableno by 1
\relax\ifproofmode\ifforwardreference
\immediate\write1{\noexpand\expandafter\noexpand\def
\noexpand\csname TABLABEL#1\endcsname{\the\chapno.\the\tableno?}}\fi\fi
\global\expandafter\edef\csname TABLABEL#1\endcsname
{\the\chapno.\the\tableno?}\iftablechapternumbers\chapfolio.\fi
\ifproofmode\llap{\hbox{\marginstyle#1
\kern1.2truein}}\relax\fi\the\tableno}

\def\tabref#1{\hbox{\ifundefined{TABLABEL#1}!!!\ifproofmode\ifforwardreference%
\else\write16{ ***Undefined Table Reference #1*** }\fi
\else\write16{ ***Undefined Table Reference #1*** }\fi
\else\edef\LABtt{\getlabel{TABLABEL#1}}%
\def\LABTT{\expandafter\stripchap\LABtt}\iftablechapternumbers%
\chapshow{\LABTT}.\expandafter\stripeq\LABtt%
\else\ifnum\number\LABTT=\chapno\relax\expandafter\stripeq\LABtt%
\else\chapshow{\LABTT}.\expandafter\stripeq\LABtt\fi\fi\fi}%
\ifproofmode\write2{Table#1}\fi}

\newdimen\sectionskip     \sectionskip=20truept
\newcount\sectno
\def\section#1#2{\sectno=0 \null\vskip\sectionskip
    \centerline{\chaplabel{#1}.~~{\bf#2}}\nobreak\vskip.2truein
    \noindent\ignorespaces}

\def\advancesectno{\global\advance\sectno by 1}
\def\sectfolio{\number\sectno}
\def\subsection#1{\goodbreak\advancesectno\null\vskip10pt
                  \noindent\chapfolio.~\sectfolio.~{\bf #1}
                  \nobreak\vskip.05truein\noindent\ignorespaces}

\def\uttg#1{\null\vskip.1truein
    \ifproofmode \line{\hfill{\bf Draft}:
    UTTG--{#1}--\number\year}\line{\hfill\today}
    \else \line{\hfill UTTG--{#1}--\number\year}
    \line{\hfill\ifcase\month\or January\or February\or March\or April%
    \or May\or June\or July\or August\or September\or October\or November%
    \or December\fi
    \space\number\year}\fi}

\def\contents{\noindent
   {\bf Contents\Z}\nobreak\vskip.05truein\noindent\ignorespaces}

\def\getlabel#1{\csname#1\endcsname}
\def\ifundefined#1{\expandafter\ifx\csname#1\endcsname\relax}
\def\stripchap#1.#2?{#1}
\def\stripeq#1.#2?{#2}

%
\catcode`@=11 
\def\space@ver#1{\let\@sf=\empty\ifmmode#1\else\ifhmode%
\edef\@sf{\spacefactor=\the\spacefactor}\unskip${}#1$\relax\fi\fi}
\newcount\referencecount     \referencecount=0
\newif\ifreferenceopen       \newwrite\referencewrite
\newtoks\rw@toks
\def\refmark#1{\relax[#1]}
\def\refend{\refmark{\number\referencecount}}
\newcount\lastrefsbegincount \lastrefsbegincount=0
\def\refsend{\refmark{\count255=\referencecount%
\advance\count255 by -\lastrefsbegincount%
\ifcase\count255 \number\referencecount%
\or\number\lastrefsbegincount,\number\referencecount%
\else\number\lastrefsbegincount-\number\referencecount\fi}}
\def\refch@ck{\chardef\rw@write=\referencewrite
\ifreferenceopen\else\referenceopentrue
\immediate\openout\referencewrite=referenc.texauxil \fi}
%
{\catcode`\^^M=\active 
  \gdef\obeyendofline{\catcode`\^^M\active \let^^M\ }}%
%
{\catcode`\^^M=\active 
  \gdef\ignoreendofline{\catcode`\^^M=5}}
{\obeyendofline\gdef\rw@start#1{\def\t@st{#1}\ifx\t@st\blankend%
\endgroup\@sf\relax\else\ifx\t@st\bl@nkend\endgroup\@sf\relax%
\else\rw@begin#1
\backtotext
\fi\fi}}
{\obeyendofline\gdef\rw@begin#1
{\def\n@xt{#1}\rw@toks={#1}\relax%
\rw@next}}
\def\blankend{}
{\obeylines\gdef\bl@nkend{
}}
\newif\iffirstrefline  \firstreflinetrue
\def\rwr@teswitch{\ifx\n@xt\blankend\let\n@xt=\rw@begin%
\else\iffirstrefline\global\firstreflinefalse%
\immediate\write\rw@write{\noexpand\obeyendofline\the\rw@toks}%
\let\n@xt=\rw@begin%
\else\ifx\n@xt\rw@@d \def\n@xt{\immediate\write\rw@write{%
\noexpand\ignoreendofline}\endgroup\@sf}%
\else\immediate\write\rw@write{\the\rw@toks}%
\let\n@xt=\rw@begin\fi\fi\fi}
\def\rw@next{\rwr@teswitch\n@xt}
\def\rw@@d{\backtotext} \let\rw@end=\relax
\let\backtotext=\relax

\newdimen\refindent     \refindent=30pt
\def\Textindent#1{\noindent\llap{#1\enspace}\ignorespaces}
\def\refitem#1{\par\hangafter=0 \hangindent=\refindent\Textindent{#1}}
\def\REFNUM#1{\space@ver{}\refch@ck\firstreflinetrue%
\global\advance\referencecount by 1 \xdef#1{\the\referencecount}}
\def\refnum#1{\space@ver{}\refch@ck\firstreflinetrue%
\global\advance\referencecount by 1\xdef#1{\the\referencecount}\refend}

\def\REF#1{\REFNUM#1%
\immediate\write\referencewrite{%
\noexpand\refitem{#1.}}%
\begingroup\obeyendofline\rw@start}
\def\ref{\refnum\?%
\immediate\write\referencewrite{\noexpand\refitem{\?.}}%
\begingroup\obeyendofline\rw@start}
\def\Ref#1{\refnum#1%
\immediate\write\referencewrite{\noexpand\refitem{#1.}}%
\begingroup\obeyendofline\rw@start}
\def\REFS#1{\REFNUM#1\global\lastrefsbegincount=\referencecount%
\immediate\write\referencewrite{\noexpand\refitem{#1.}}%
\begingroup\obeyendofline\rw@start}

\def\REFSCON#1{\REF#1}

\def\cite#1{\refmark#1}
\catcode`@=12 
%
%
%
%
\expandafter \def \csname CHAPLABELintro\endcsname {1}
\expandafter \def \csname CHAPLABELmanifold\endcsname {2}
\expandafter \def \csname CHAPLABELmoduli1\endcsname {3}
\expandafter \def \csname EQLABELyeqs\endcsname {3.1?}
\expandafter \def \csname EQLABELxeqs\endcsname {3.2?}
\expandafter \def \csname CHAPLABELnew\endcsname {4}
\expandafter \def \csname EQLABELlcsl\endcsname {4.1?}
\expandafter \def \csname EQLABELmms\endcsname {4.2?}
\expandafter \def \csname EQLABELjordan\endcsname {4.3?}
\expandafter \def \csname CHAPLABELmore\endcsname {5}
\expandafter \def \csname EQLABELfamily\endcsname {5.1?}
\expandafter \def \csname EQLABELedges\endcsname {5.2?}
\expandafter \def \csname TABLABELfourregs1\endcsname {5.1?}
\expandafter \def \csname TABLABELfourregs2\endcsname {5.2?}
\expandafter \def \csname CHAPLABELfundper\endcsname {6}
\expandafter \def \csname EQLABELpi0double\endcsname {6.1?}
\expandafter \def \csname EQLABELpi12\endcsname {6.2?}
\expandafter \def \csname EQLABELPFeqns\endcsname {6.3?}
\expandafter \def \csname EQLABELpi0single\endcsname {6.4?}
\expandafter \def \csname EQLABELydef\endcsname {6.5?}
\expandafter \def \csname EQLABELhgfs\endcsname {6.6?}
\expandafter \def \csname EQLABELintrep\endcsname {6.7?}
\expandafter \def \csname EQLABELintegers\endcsname {6.8?}
\expandafter \def \csname EQLABELderiv\endcsname {6.9?}
\expandafter \def \csname EQLABELbigpsi\endcsname {6.10?}
\expandafter \def \csname EQLABELmono\endcsname {6.11?}
\expandafter \def \csname EQLABELvn\endcsname {6.12?}
\expandafter \def \csname EQLABELveq\endcsname {6.13?}
\expandafter \def \csname EQLABELdiffnu\endcsname {6.14?}
\expandafter \def \csname EQLABELweq\endcsname {6.15?}
\expandafter \def \csname EQLABELrecur\endcsname {6.16?}
\expandafter \def \csname TABLABELpolys\endcsname {6.1?}
\expandafter \def \csname EQLABELpu12\endcsname {6.17?}
\expandafter \def \csname EQLABELpismall\endcsname {6.18?}
\expandafter \def \csname EQLABELmnint\endcsname {6.19?}
\expandafter \def \csname EQLABELmunu\endcsname {6.20?}
\expandafter \def \csname EQLABELxieta\endcsname {6.21?}
\expandafter \def \csname CHAPLABELflatcoords\endcsname {7}
\expandafter \def \csname EQLABELPidef\endcsname {7.1?}
\expandafter \def \csname EQLABELipdef\endcsname {7.2?}
\expandafter \def \csname EQLABELflatcds\endcsname {7.3?}
\expandafter \def \csname EQLABELlcslbis\endcsname {7.4?}
\expandafter \def \csname EQLABELmmap\endcsname {7.5?}
\expandafter \def \csname EQLABELprepotF\endcsname {7.6?}
\expandafter \def \csname EQLABELSp\endcsname {7.7?}
\expandafter \def \csname EQLABELident\endcsname {7.8?}
\expandafter \def \csname EQLABELmdef\endcsname {7.9?}
\expandafter \def \csname EQLABELmconds\endcsname {7.10?}
\expandafter \def \csname EQLABELts\endcsname {7.11?}
\expandafter \def \csname EQLABELtseqs\endcsname {7.12?}
\expandafter \def \csname EQLABELiter\endcsname {7.13?}
\expandafter \def \csname EQLABELuhfhat\endcsname {7.14?}
\expandafter \def \csname CHAPLABELyukawas\endcsname {8}
\expandafter \def \csname EQLABELyper\endcsname {8.1?}
\expandafter \def \csname EQLABELtangsp\endcsname {8.2?}
\expandafter \def \csname EQLABELinstantons\endcsname {8.3?}
\expandafter \def \csname EQLABELcot\endcsname {8.4?}
\expandafter \def \csname TABLABELinsts\endcsname {8.1?}
\expandafter \def \csname EQLABELinstantons2\endcsname {8.5?}
\expandafter \def \csname EQLABELcijk\endcsname {8.6?}
\expandafter \def \csname TABLABELinst12\endcsname {8.2?}
\expandafter \def \csname EQLABELtopF\endcsname {8.7?}
\expandafter \def \csname EQLABELFinst\endcsname {8.8?}
\expandafter \def \csname EQLABELfhol\endcsname {8.9?}
\expandafter \def \csname EQLABELlimF\endcsname {8.10?}
\expandafter \def \csname TABLABELdinsts\endcsname {8.3?}
\expandafter \def \csname TABLABELdinst12\endcsname {8.4?}
\expandafter \def \csname CHAPLABELverification\endcsname {9}
\expandafter \def \csname EQLABEL3pt\endcsname {9.1?}
%
\def\hourandminute{\count255=\time\divide\count255 by 60
\xdef\hour{\number\count255}
\multiply\count255 by -60\advance\count255 by\time
\hour:\ifnum\count255<10 0\fi\the\count255}

\def\subsection#1{\goodbreak\advancesectno\null\vskip10pt
                  \noindent{\it \chapfolio.\sectfolio.~#1}
                  \nobreak\vskip.05truein\noindent\ignorespaces}

\def\cite#1{\refmark{#1}}

\def\:{\kern-1.5truept}

\def\bar{\overline}

\def\\{\hfill\break}

\def\cropen#1{\crcr\noalign{\vskip #1}}

\def\yzero{\smash{\hbox{$y\kern-4pt\raise1pt\hbox{${}^\circ$}$}}}

\def\contents{\line{{\bf Contents}\hfill}\nobreak\vskip.05truein\noindent%
              \ignorespaces}

\def\F#1#2#3#4{{}_2 F_1\left({#1},\,{#2};\,{#3};\,{#4}\right)}

\def\scr{\scriptstyle}

\def\txt{\textstyle}

\def\ds{\displaystyle}

\def\crs{\cr\noalign{\smallskip}}

\def\crm{\cr\noalign{\medskip}}

\def\crb{\cr\noalign{\bigskip}}

\def\point#1{\noindent\setbox0=\hbox{#1}\kern-\wd0\box0}

\def\unu{u_\n(\ph)}

\def\unum{u_\n(-\ph)}

\def\vnu{v_\n(\ph)}

\def\vn{v_n(\ph)}

\def\un{u_n(\ph)}

\def\unm{u_n(-\ph)}

\def\wn{w_n(\ph)}

\def\fn{f_n(\ph)}

\def\gn{g_n(\ph)}

\def\lcsl{large complex structure limit}

\def\Mone{\ifmmode \cp4^{(1,1,2,2,2)}[8]\else $\cp4^{(1,1,2,2,2)}[8]$\fi}

\def\Mtwo{\ifmmode \cp4^{(1,1,2,2,6)}[12]\else $\cp4^{(1,1,2,2,6)}[12]$\fi}

\def\wP{\hbox{$\IP^{(1,1,2,2,2)}$}}
\def\WP{\hbox{$\IP^{(1,1,2,2,6)}$}}
\def\Hom{{\rm Hom}}
\def\barT{\hbox{$\bar T$}}
\def\barS{\hbox{$\bar S$}}

\def\barN{\hbox{$\bar N$}}

\def\sA{\hbox{\ss A}}
\def\sB{\hbox{\ss B}}
\def\sC{\hbox{\ss C}}
\def\sD{\hbox{\ss D}}

\def\sR{\hbox{\ss R}}
\def\sS{\hbox{\ss S}}
\def\sT{\hbox{\ss T}}
\def\sY{\hbox{\ss Y}}
\def\T{\ca{T}}

\def\W{\ca{W}}

\def\cS{\ca{S}}


\def\matrixa{\pmatrix{\- 0 &\- 1 &\- 0 &\- 0 &\- 0 &\- 0 \cr
                      \- 0 &\- 0 &\- 1 &\- 0 &\- 0 &\- 0 \cr
                      \- 0 &\- 0 &\- 0 &\- 1 &\- 0 &\- 0 \cr
                      \- 0 &\- 0 &\- 0 &\- 0 &\- 1 &\- 0 \cr
                      \- 0 &\- 0 &\- 0 &\- 0 &\- 0 &\- 1 \cr
                        -1 &\- 0 &  -1 &\- 0 &  -1 &\- 0 \cr  }}

\def\matrixb{\pmatrix{\- 1 &\- 0 &\- 0 &\- 0 &\- 0 &\- 0 \cr
                      \- 0 &\- 1 &  -1 &\- 1 &  -1 &\- 1 \cr
                      \- 0 &\- 0 &\- 2 &  -1 &\- 1 &  -1 \cr
                      \- 3 &  -3 &\- 4 &  -3 &\- 3 &  -3 \cr
                        -3 &\- 3 &  -4 &\- 4 &  -2 &\- 3 \cr
                        -3 &\- 3 &  -3 &\- 3 &  -2 &\- 3 \cr  }}

\def\matrixt{\pmatrix{\- 2 &  -1 &\- 0 &\- 0 &\- 0 &\- 0 \cr
                      \- 1 &\- 0 &\- 0 &\- 0 &\- 0 &\- 0 \cr
                        -1 &\- 1 &\- 1 &\- 0 &\- 0 &\- 0 \cr
                        -3 &\- 3 &\- 0 &\- 1 &\- 0 &\- 0 \cr
                      \- 3 &  -3 &\- 0 &\- 0 &\- 1 &\- 0 \cr
                      \- 3 &  -3 &\- 0 &\- 0 &\- 0 &\- 1 \cr  }}

\def\matrixtinf{\pmatrix{\- 1&\- 0&\- 0&\- 0&\- 0&\- 0\cr
                         \- 2&\- 1&\- 0&\- 1&\- 0&\- 1\cr
                           -1&\- 0&\- 0&  -1&\- 0&  -1\cr
                           -3&  -3&\- 1&  -3&\- 0&  -3\cr
                         \- 3&\- 3&\- 0&\- 4&\- 0&\- 3\cr
                         \- 3&\- 3&\- 0&\- 3&\- 1&\- 3\cr}}

\def\matrixTF{\pmatrix{\- 1 &\- 0 &\- 0 &\- 0 &\- 0 &\- 0 \cr
                       \- 0 &\- 1 &\- 0 &\- 0 &\- 0 &\- 0 \cr
                       \- 0 &\- 0 &\- 1 &\- 0 &\- 0 &\- 0 \cr
                         -1 &\- 0 &\- 0 &\- 1 &\- 0 &\- 0 \cr
                       \- 0 &\- 0 &\- 0 &\- 0 &\- 1 &\- 0 \cr
                       \- 0 &\- 0 &\- 0 &\- 0 &\- 0 &\- 1 \cr }}

\proofmodefalse
\baselineskip=13pt
\parskip=2pt
\chapternumberstrue
\figurechapternumberstrue
\tablechapternumberstrue
\forwardreferencefalse
\ifproofmode
\immediate\openout2=allcrossreferfile \fi
\ifforwardreference\input labelfile
\ifproofmode\immediate\openout1=labelfile \fi\fi
\noblackboxes
\hfuzz=1pt
\vfuzz=2pt
%
\nopagenumbers\pageno=0
\null\vskip-50pt
\rightline{\eightrm CERN-TH.~6884/93, UTTG-15-93}\vskip-3pt
\rightline{\eightrm NEIP-93-005, OSU-M-93-1}\vskip-3pt
\rightline{\eightrm hep-th/9308083}\vskip-3pt
\rightline{\eightrm August 25, 1993}
\vskip.2truein
\centerline{\seventeenrm \hphantom{\ \raise6pt\hbox{*}}Mirror Symmetry
for Two Parameter Models -- I\ \raise6pt\hbox{*}}
\vfootnote{\eightrm *}{\eightrm  Supported in part by the Robert A. Welch
                       Foundation, NSF grants PHY-9009850 and DMS-9103827,
                       the Swiss National Foundation, and an American
                       Mathematical Society Centennial Fellowship.
}
\vskip.3truein
\centerline{%
      {\csc Philip~Candelas}$^{1,2}$,\quad
      {\csc Xenia~de la Ossa}$^3$,\quad
      {\csc Anamar\'\i a~Font}$^4$,}
\vskip1mm
\centerline{{\csc Sheldon Katz}$^5$ {\csc and David R.~Morrison}$^{6,7}$}
\vskip.2truein\bigskip
\centerline{
\vtop{\hsize = 2.0truein
\centerline {$^1$\it Theory Division}
\centerline {\it  CERN}
\centerline {\it CH-1211 Geneva 23}
\centerline {\it Switzerland}}
\vtop{\hsize = 2.0truein
\centerline{$^2$\it Theory Group}
\centerline{\it Department of Physics}
\centerline{\it University of Texas}
\centerline{\it Austin, TX 78712, USA}}
\vtop{\hsize = 2.0truein
\centerline{$^3$\it Institut de Physique}
\centerline{\it Universit\'e\:\ de\:\ Neuch\^atel}
\centerline{\it CH-2000 Neuch\^atel}
\centerline{\it Switzerland}}}
\vskip.2truein
\centerline{
\vtop{\hsize = 3.0truein
\centerline{$^4$\it Departamento de F\'\i sica}
\centerline{\it Universidad Central de Venezuela}
\centerline{\it A.P.~20513,\:\ Caracas 1020--A}
\centerline{\it Venezuela}}
\hskip0pt
\vtop{\hsize = 3.0truein
\centerline{$^5$\it Department of Mathematics}
\centerline{\it Oklahoma State University}
\centerline{\it Stillwater}
\centerline{\it OK 74078, USA}}}
\vskip.2truein
\centerline{
\vtop{\hsize = 3.0truein
\centerline{$^6$\it School of Mathematics}
\centerline{\it Institute for Advanced Study}
\centerline{\it Princeton, NJ 08540, USA}}
\hskip0pt
\vtop{\hsize = 3.0truein
\centerline{$^7$\it Department of Mathematics}
\centerline{\it Duke University}
\centerline{\it Durham, NC 27708, USA}}}
\bigskip\bigskip\bigskip
\nobreak\vbox{\centerline{\bf ABSTRACT}
\vskip.25truein
\vbox{\baselineskip 11.5pt\noindent We study, by means of mirror symmetry, the
quantum geometry of the \K-class parameters of a number of Calabi--Yau
manifolds that have $b_{11}=2$. Our main interest lies in the structure of the
moduli space and in the loci corresponding to singular models. This structure
is considerably richer when there are two parameters than in the various
one-parameter models that have been studied hitherto. We describe the intrinsic
structure of the point in the (compactification of the) moduli space that
corresponds to the large complex structure or classical limit. The instanton
expansions are of interest owing to the fact that some of the instantons belong
to families with continuous parameters. We compute the Yukawa couplings and
their expansions in terms of instantons of genus zero. By making use of recent
results of Bershadsky {\it et al.\/} we compute also the instanton numbers for
instantons of genus one. For particular values of the parameters the models
become birational to certain models with one parameter. The compactification
divisor of the moduli space thus contains copies of the moduli spaces of one
parameter models. Our discussion proceeds via the particular models
\smash{$\cp4^{(1,1,2,2,2)}[8]$ and $\cp4^{(1,1,2,2,6)}[12]$}. Another example,
\smash{$\cp4^{(1,1,1,6,9)}[18]$}, that is somewhat different is the subject of
a companion~paper.
\smallskip
\leftline{\eightrm CERN-TH.~6884/93}}}
\newpage
\headline={\ifproofmode\hfil\eightrm draft:\ \today\
\hourandminute\else\hfil\fi}

\pageno=0
\contents
\vskip15pt
\line{1.~~Introduction\hfil}\bigskip
\line{2.~~Geometry of \cy\ Hypersurfaces in $\IP^{(1,1,2,2,2)}$
          and $\IP^{(1,1,2,2,6)}$\hfil}
\itemitem{\it 2.1~}{\it Linear systems}
\itemitem{\it 2.2~}{\it Curves and the K\"ahler cone}
\itemitem{\it 2.3~}{\it Chern classes}\bigskip
\line{3.~~The Moduli Space of the Mirror\hfil}
\itemitem{\it 3.1~}{\it Basic facts}
\itemitem{\it 3.2~}{\it The locus $\ph^2=1$}
\itemitem{\it 3.2~}{\it More about the moduli space}\bigskip
\line{4.~~Monodromy and the Large Complex Structure Limit\hfil}
\itemitem{\it 4.1~}{\it The large complex structure limit}
\itemitem{\it 4.2~}{\it Monodromy calculations}\bigskip
\line{5.~~Considerations of Toric Geometry\hfil}\bigskip
\line{6.~~The Periods\hfil}
\itemitem{\it 6.1~}{\it The fundamental period}
\itemitem{\it 6.2~}{\it The Picard--Fuchs equations}
\itemitem{\it 6.3~}{\it Analytic properties of the fundamental period}
\itemitem{\it 6.4~}{\it Analytic continuation of the periods}\bigskip
\line{7.~~The Mirror Map and the Large Complex Structure Limit\hfil}
\itemitem{\it 7.1~}{\it Generalities}
\itemitem{\it 7.2~}{\it The large complex structure limit for \Mone}
\itemitem{\it 7.3~}{\it Inversion of the mirror map}\bigskip
\line{8.~~The Yukawa Couplings and the Instanton Expansion\hfil}
\itemitem{\it 8.1~}{\it The couplings}
\itemitem{\it 8.2~}{\it Instantons of genus one}\bigskip
\line{9.~~Verification of Some Instanton Contributions\hfil}\bigskip

\newpage
\footline={\rm\hfil\folio\hfil}
\section{intro}{Introduction}
In this article we study the mirror map for the moduli spaces of two
two-parameter \cys. This study draws on and extends the methods of~
\REFS\rCdGP{P.~Candelas, X.~ de la Ossa, P.~Green and L.~Parkes,
       \npb{359} (1991) 21.}
\REFSCON\rMorrison{D.~R.~Morrison, ``Picard-Fuchs Equations and Mirror Maps
       for Hypersurfaces'', in {\it Essays on Mirror Symmetry}, ed.\
       S.-T.~Yau (Intl. Press, Hong Kong, 1992).}
\REFSCON\rFont{A.~Font, \npb{391} (1993) 358.}
\REFSCON\rKT{A.~Klemm and S.~Theisen, \npb{389} (1993) 153.}
\refsend.
We have undertaken this investigation because the multiparameter case is
generic and is considerably more involved than the one parameter models that
have been studied hitherto. We are thus driven to methods of more general
applicability.
Furthermore the structure of the moduli space is considerably richer in the two
parameter case. The compactification of the moduli space contains curves
corresponding to several types of singularities and we address the interesting
question of how to characterize and identify the point in the moduli space
corresponding to the large complex structure limit. We are led also to
investigate the analytic properties of the periods as functions of two
parameters. Another interesting feature is that the \cys\ that we investigate
have instantons which lie in continuous families for generic values of the
parameters. One consequence of this is that the instantons belonging to the
continuous family cannot be completely separated from the classical
contribution to the Yukawa couplings. Another interesting feature of the
instanton expansion is that we find certain `quantum symmetries' pertaining to
the instanton numbers. These symmetries may be understood in virtue of certain
monodromy transformations of the period vector for the complex structure
parameters of the mirror manifold. They can also be verified and
generalized by the methods of
algebraic geometry, but these results would not have been expected
without mirror symmetry.
We find also that certain instanton numbers are negative
contrary, perhaps, to naive expectation though since these numbers correspond
to integrals over continuous families of instantons there is in reality no
contradiction.

The hardest part of the analysis concerns the proper identification of the
large complex structure limit and the explicit construction of the mirror map.
The methods previously applied to one parameter models can be employed in the
present context also. The multi-parameter case is however considerably more
involved than the one parameter case and a clearer and more general procedure
is desirable and is developed here. We show how the \lcsl\ is characterised by
certain monodromy properties and we show also, by means of an exhaustive
calculation of monodromy transformations, that the point corresponding to the
\lcsl\ is uniquely determined. We also verify that a simpler procedure based on
the methods of toric geometry (pioneered in~
\REFS\rAGM{P.~S.~Aspinwall, B.~R.~Greene and D.~R.~Morrison,
 \plb{303} (1993) 249;
 ``Calabi-Yau Moduli Space, Mirror Manifolds, and Spacetime Topology
 Change in String Theory'', in preparation.}
\refsend)
produces the same result.  For the construction
of the mirror map itself, we analyze the monodromy more carefully using
analytic continuation of periods.

Our study proceeds as follows: we explain in \SS2 the geometry of our two \cys\
and the structure of their \K-class moduli spaces. In \SS3 we discuss the
structure of the complex structure moduli spaces for their mirrors. In virtue
of mirror\vadjust{\newpage}
symmetry these are the quantum corrected versions of the moduli
spaces of \SS2. We turn in \SS4 to a discussion of the monodromy properties
that characterise the \lcsl. These monodromy properties relate to the transport
of the periods about the various curves in the moduli space that correspond to
singular degenerations of the \cym. Singularities in the \cym\ produce
metric singularities in the moduli space.
To properly analyze these, we need to compactify the moduli space to
include a normal crossings divisor on the boundary.
This
leads us to consider, in \SS5, certain techniques of toric geometry which
provide another, perhaps faster, way of finding the \lcsl. We turn in \SS6 to a
computation of the periods for our models. These are certain generalized
hypergeometric functions of two variables and we are concerned with the related
issues of their analytic continuation throughout the moduli space and their
monodromy about the curves in the moduli space that correspond to singular
manifolds. In \SS7 we apply the foregoing analysis to finding the mirror map
between the flat coordinates on the \K-cone of the models and the complex
structure parameters of their mirrors and in \SS8 we compute the Yukawa
couplings and exhibit their expansion in terms of instantons. Since we have the
periods and the various expansions to hand we compute also the instanton
expansion of a certain genus-one generalization of the prepotential introduced
by Bershadsky {\it et al.\/}~
\Ref\rBCOV{M.~Bershadsky, S.~Cecotti, H.~Ooguri and C.~Vafa,
        ``Holomorphic Anomalies in Topological Field Theories'', with
          an appendix by S.~Katz, Harvard
          University preprint HUTP-93/A008, RIMS-915.}\
and obtain also the instanton numbers for instantons of genus one. Finally in
\SS9 we check the values of some of the instanton numbers that we have found,
and explore the meaning of the negativity of some of those numbers.

After this work was complete, we were kindly informed by S.-T.~Yau
of a preprint~
\Ref{\rHKTY}{S.~Hosono, A.~Klemm, S.~Theisen, and S.-T.~Yau,
``Mirror Symmetry, Mirror Map, and Application to Calabi-Yau Hypersurfaces'',
Preprint.}\
which overlaps with the present paper.
\newpage
\section{manifold}{Geometry of \cy\ Hypersurfaces in $\IP^{(1,1,2,2,2)}$
and $\IP^{(1,1,2,2,6)}$}
\vskip-20pt
\subsection{Linear systems}
We consider Calabi-Yau threefolds $\ca{M}$ which are obtained by resolving
singularities of degree eight hypersurfaces
$\Mhat\subset\IP^{(1,1,2,2,2)}$.
A typical defining polynomial for such a hypersurface is
$$p=x_1^8 + x_2^8 + x_3^4 + x_4^4 + x_5^4 $$
but of course in general many other terms can occur.

The singularities occur along $x_1=x_2=0$, where there is a curve $C$ of
singularities of type $A_1$.  In our particular example,
the curve $C$ is described by
$$
x_1=x_2=0~~~,~~~x_3^4 + x_4^4 + x_5^4=0~;$$
in general it will just
be a smooth quartic plane curve, which always has genus $3$.

To resolve singularities, we must blow up the locus $x_1=x_2=0$; the
curve of singularities $C$ on $\Mhat$ is replaced by a divisor $E$
on $\ca{M}$, which is a ruled surface over the curve $C$.  That is
each point of $C$ is blown up into a \cp{1}.

Recall that on a complex manifold, a (complete) {\sl linear system}
is the set of divisors arising as the zero loci of the global sections
of a line bundle.  Two divisors in the same linear system are said to be
linearly equivalent; and linearly equivalent divisors may be freely
substituted for each other in an intersection calculation.  Given an
effective divisor $D$, the complete linear system of which it is a part is
denoted by $|D|$.

The first linear system we study on $\ca{M}$, which we will denote by
$|L|$,
is generated by polynomials of degree 1 (i.e., by $x_1$ and $x_2$).
Every divisor in $|L|$ is the proper transform on $\ca{M}$ of the zero
locus
of such a polynomial on $\Mhat$.  These divisors are described by
means of a parameter $\lambda$ and a substitution $x_2=\lambda x_1$
($\lambda$ may be infinite); the equation of the proper transform
becomes
$$
(1+\lambda^8)x_1^8+x_3^4 + x_4^4 + x_5^4=0$$
which defines a surface of  degree 8 in $\IP_3^{(1,2,2,2)}$.
In fact, making the substitution $y_1=x_1^2$, which is single valued
in virtue of the scaling properties of the coordinates, we see that this is
isomorphic to the surface of degree 4 in ordinary projective space
with equation
$$
(1+\lambda^8)y_1^4+x_3^4 + x_4^4 + x_5^4=0.$$
The linear system $|L|$ is thus a pencil of quartic K3 surfaces.
Note that any two distinct members of $|L|$ are disjoint,
that is, $L\cdot L=0$.

The second linear system we study on $\ca{M}$ (which we will denote by
$|H|$)
is generated by polynomials of degree 2 (i.e., by linear combinations of
$x_1^2$, $x_1x_2$, $x_2^2$, $x_3$, $x_4$, and $x_5$).
The divisors in $|H|$ are total transforms on $\ca{M}$ of the zero locus
on $\Mhat$ of the corresponding polynomial.  A typical polynomial
will have non-zero coefficient on $x_5$, and allow one to solve for
$x_5$ in terms of the other variables, producing a proper transformed
equation which defines a surface of degree 8 in $\IP_3^{(1,1,2,2)}$.

These two linear systems are related to each other as follows.  If we
look at $|2L|$, the quadratic polynomials in $x_1$ and $x_2$, we get
a subsytem of $H$.  That subsytem can be characterized by the geometric
property that the polynomials from $|2L|$ vanish on the singular curve
$C$.  Interpreted on the resolution $\ca{M}$, this means that the {\sl
total}
transform of the zero locus of such a polynomial has the form $2L+E$
where $2L$ describes the proper transform and $E$ is the exceptional
divisor (as above).  We thus arrive at the relation between linear systems
$$|H|=|2L+E|.$$

We now compute intersection products.  Since $L\cdot L=0$, we automatically
have
$$
H\cdot L^2=0~,~~~~~L^3=0~.$$
Since $|H|$ defines a birational map on $\ca{M}$ whose image has degree 8
(the number of common intersection points of three members of
$|H|$), we have\Footnote{Typically in such an intersection calculation
we express $x_3$, $x_4$, and $x_5$ as quadratic polynomials in $x_1$
and $x_2$, leaving us with an overall homogeneous octic in 2 variables,
which will have 8 solutions.}
$$
H^3=8.$$
And when we restrict that linear system $|H|$ to one of the K3 surfaces
$L$, we get a {\sl quartic} linear system on $L$.  It follows
that\Footnote{In this
calculation, we have $x_2=\lambda x_1$, so we express $x_4$ and $x_5$
as linear polynomials in $x_1^2$ and $x_3$, that is, in $y_1$ and $x_3$.
The equation becomes a homoegeneous quartic in $y_1$ and $x_3$, giving
4 intersection points.}
$$
H^2\cdot L=(H\cap L)\cdot(H\cap L)=4.$$

The situation is entirely analogous for degree 12 hypersurfaces
$\Mhat\subset\IP^{(1,1,2,2,6)}$.  A~typical defining polynomial is
$$
p=x_1^{12}+x_2^{12}+x_3^6+x_4^6+x_5^2~.$$
The singular curve $C$ is described by
$$
x_1=x_2=0~~~,~~~x_3^6 + x_4^6 + x_5^2=0~;$$
which is a curve of genus 2.  The singularity is resolved by blowing up
$C$, resulting in an exceptional divisor $E$ which is a ruled surface over
the curve $C$.  The degree 1 polynomials generated by $x_1$ and $x_2$ define
a linear system $|L|$, whose divisors are again K3 surfaces (this time
in $\IP^{(1,2,2,6)}$, and are isomorphic to surfaces of degree 6 in
$\IP^{(1,1,1,3)}$).  The degree 2 polynomials generate a linear system
$|H|$.  We again have $H=2L+E$.  The intersection numbers are
$$
H^3=4~,\qquad H^2L=2~,\qquad HL^2=0~,\qquad L^3=0~.$$
In fact, $L^2=0$, as in the previous case.
\subsection{Curves and the K\"ahler cone}
We now consider the classes of some $1$-cycles on
$\ca{M}\subset\IP^{(1,1,2,2,2)}$.  Our first
class
is $l$, a fiber of the ruling $E\to C$.  We identify its cohomology class
by noting that $H\cap E$ consists of 4 fibers (lying over the 4 points
of intersection of the hyperplane with $C$), so that
$$l=\frac14H\cdot E=\frac14H^2-\frac12H\cdot L.$$

The second class we consider is the intersection of general members of
$|H|$
and $|L|$,
$$4h=H\cdot L.$$
This is a plane quartic curve (after the substitution
$y_1 = x_1^2$ has been made).  In fact, by choosing H and L carefully, we
can
find a curve which is a sum of four lines. An easy way to do this is to
take
$L$ to be $x_2 = 0$, and to take $H$ to be $x_5 = e^{2\pi i/8} x_4$. Then
each
of the lines defines a class $h$.

The intersection relations between linear systems and curves read
$$\eqalign{
L\cdot l&=1\cr
H\cdot l&=0\cr}\hskip30pt
\eqalign{
L\cdot h&=0\cr
H\cdot h&=1~.\cr}$$

\bigskip

We can now identify the K\"ahler cone of our variety $\ca{M}$.  If we take
a
general linear system $|\alpha L+\beta H|$, by intersecting with
the effective classes $l$ and $h$ we get constraints on $\alpha$
and $\beta$.  From the intersection relations given above, we see that
those
constraints take the particularly simple form $\alpha\ge0$, $\beta\ge0$.

On the other hand, the extreme solutions to those constraints, namely
$L$ and $H$, are known to represent classes of nef divisors, since
the linear systems $|L|$ and $|H|$ are base-point-free on $\ca{M}$.  They
are thus limits of sequences of K\"ahler classes; we conclude that
the K\"ahler cone is precisely the convex hull of $\IR_{\ge 0}L$ and
$\IR_{\ge 0}H$.

The same discussion applies for $\Mhat\subset\IP^{(1,1,2,2,6)}$, with
the obvious modifications.  The result is the same: the K\"ahler
cone is the convex hull of $\IR_{\ge 0}L$ and
$\IR_{\ge 0}H$.

Both $\Mone$ and $\Mtwo$ have non-polynomial deformations, but by results
in~
\REFS\rWilson{P.M.H.~Wilson, Invent.\ Math.\ {\bf 107} (1992) 561}
\refsend,
their K\"ahler cones will also be the convex hull of
$\IR_{\ge 0}L$ and $\IR_{\ge 0}H$.

\subsection{Chern classes}
The second Chern class can be computed without too much difficulty and will be
of use later in relation to the instantons of genus one.
For a smooth divisor $D\subset\ca{M}$, we use the notation $c_2(D)$
to note the second Chern class of the surface $D$.  The notation
$c_i$ will be reserved for the appropriate Chern class of $\ca{M}$.
For the K3 surface $L$, we have $c_2(L)=24$.  And for the exceptional
divisor $E$, a ruled surface over a curve of genus $g$ ($g$ is 3 or 2 in
our respective cases), we have
$c_2(E)=4-4g$.

Let $T_D,\ T_{\ca{M}}$ respectively denote the holomorphic tangent bundles of
$D,\ \ca{M}$, and let $N_{D/\ca{M}}$ denote the normal bundle of $D$ in
$\ca{M}$; note
$$
N_{D/\ca{M}}\simeq\ca{O}(D)|_D.$$
{}From the exact sequence
$$
0\rightarrow T_D\rightarrow T_{\ca{M}}|_D\to N_{D/\ca{M}}\to 0$$
together with the Whitney product formula, we calculate
$$
(1+c_2+c_3)|_D=((1-D+c_2(D))\cdot (1+D))_D,$$
where the subscript indicates that the intersection is to be performed on
$D$.  Using this and the fact that
$$
(D\cdot D)_D=(D\cdot D\cdot D)_{\ca{M}},$$
we obtain the desired formula
$$
c_2\cdot D=c_2(D)-D^3.$$
In our situation, this gives
$$
c_2\cdot L=c_2(L)-L^3=24.$$

For the case of $\Mone$, we have
$$
c_2\cdot E=c_2(E)-E^3=-8-(H-2L)^3=8~,$$
and hence
$$
c_2\cdot H=c_2\cdot(2L+E)=2(24)+8=56~.$$

While for $\Mtwo$ we similarly obtain
$$
c_2\cdot E=c_2(E)-E^3=-4-(H-2L)^3=4~,\qquad c_2\cdot H=52~.$$
\newpage
\section{moduli1}{The Moduli Space of the Mirror}
\vskip-20pt
\subsection{Basic facts}
Using the construction of~
\REFS{\rGP}{B.R. Greene and M.R. Plesser, Nucl. Phys. {\bf B338} (1990)
15.}
\refsend,
the mirror of $\M_1=\Mone$ may be identified with the family of Calabi-Yau
threefolds of the form $\{p=0\}/G$, where
$$p=x_1^8 + x_2^8 + x_3^4 + x_4^4 + x_5^4 - 8\psi\, x_1 x_2 x_3 x_4 x_5
                                          - 2\phi\, x_1^4 x_2^4 $$
and where $G$, which is abstractly $\IZ_4^3$, is the  group with
generators
$$\eqalign{
&(\IZ_4;~0,3,1,0,0)~,\cr
&(\IZ_4;~0,3,0,1,0)~,\cr
&(\IZ_4;~0,3,0,0,1)~.\cr}$$
For a good description of the moduli
space, it is wise to enlarge $G$ to $\widehat{G}$ consisting of
elements $g=(\a^{a_1},\a^{a_2},\a^{2a_3},\a^{2a_4},\a^{2a_5})$
acting as:
$$(x_1,x_2,x_3,x_4,x_5;\psi,\phi)\mapsto
(\a^{a_1}x_1,\a^{a_2}x_2,\a^{2a_3}x_3,\a^{2a_4}x_4,\a^{2a_5}x_5;
\a^{-a}\psi,\a^{-4a}\phi),$$
where $a=a_1+a_2+2a_3+2a_4+2a_5$,
where $\a^{a_1}$ and $\a^{a_2}$ are 8th roots of unity, and where
$\a^{2a_3}$, $\a^{2a_4}$, and $\a^{2a_5}$ are 4th roots of unity.
(We do not require that the product of these be 1, since we have `corrected'
the equation by an appropriate action on the coefficients.)

If we mod the family of weighted projective hypersurfaces $\{p=0\}$ by
the full group $\widehat{G}$, we must mod the parameter space
$\{(\psi,\phi)\}$ by a $\IZ_8$ whose generator $g_0$ acts by
$$
(\psi,\phi)\mapsto(\a\psi,-\phi).$$
The quotiented parameter space has a singularity at the origin, and
can be described by three functions
$$
\tilde{\x}\define\psi^8~,\ \ \ \tilde{\eta}\define\psi^4\phi~,\ \ \
\tilde{\z}\define\phi^2~,$$
subject to the relation
$$
\tilde{\x}\,\tilde{\z}=\tilde{\eta}^2~.$$
This describes an affine quadric cone in $\IC^3$.  When we need a
compactification of the moduli space, it will be natural to compactify
it to the projective quadric cone in $\IP_3$.
We embed $\IC^3$ in $\IP_3$ by sending the point of $\IC^3$ with
coordinates $(\tilde{\x},\tilde{\eta},\tilde{\z})$ to the point in
$\IP_3$ with {\sl homogeneous} coordinates
$[\tilde{\x},\tilde{\eta},\tilde{\z},1]$.  (The most
general point in $\IP_3$ has homogeneous coordinates
$[\x,\eta,\z,\tau]$; when $\tau\ne0$, these are related to the affine
coordinates of $\IC^3$ by
$\tilde{\x}=\x/\tau$,
$\tilde{\eta}=\eta/\tau$,
$\tilde{\z}=\z/\tau$.)

One important thing to keep in mind about this quotient by $\IZ_8$:  the
square of the generator $g_0^2$ acts trivially on $\phi$, and so fixes
the entire line $\psi=0$.  This means that the quotiented family
$\{p=0\}/\widehat{G}$ will have new singularities along the $\psi=0$
locus (which becomes the $\x=\eta=0$ locus in the quotient).

\bigskip

We need to locate the parameter values for which the original family
of hypersurfaces $\{p=0\}$ is singular, and study the behavior of the
singularities under the quotienting.  This is straightforward and produces the
following result:

\item{1.~~}
Along the locus $\phi+8\psi^4=\pm1$, the threefold $\{p=0\}$ acquires
a collection of conifold points, which are identified under the $G$-action,
giving only one conifold point per threefold on the quotient.

\item{2.~~}
Along the locus $\phi=\pm 1$, the threefold $\{p=0\}$ acquires $4$ somewhat
complicated isolated singularities, again leading to only a single
singular point on the quotient.

\item{3.~~}
If we let $\phi$ and $\psi$ approach infinity, we get a singular
limiting threefold which may for example have
 the form
$$
(8\psi\, x_1 x_2 x_3 x_4 x_5+ 2\phi\, x_1^4 x_2^4=0)/G.$$
(In this example, the limiting threefold splits into at least three components,
and must be singular along
their intersections---two of the components being $x_1=0$ and $x_2=0$.)

\item{4.~~}
As noted above, the locus $\psi=0$ requires special treatment, as it
leads to additional singularities on the quotient by $\widehat{G}$.

\noindent When we pass to the quotiented parameter space, these loci can be
described as follows.  We use homogeneous coordinates $[\x,\eta,\z,\t]$
on $\IP_3$, and describe the compactified quotiented
parameter space as the singular quadric
$Q\define\{\x\z-\eta^2=0\}\subset \IP_3$.
The loci above then become the curves:

\item{1.~~}
$C_{\rm con}=Q\cap\{64\x+16\eta+\z-\t=0\}$,

\item{2.~~}
$C_1=Q\cap\{\z-\t=0\}$,

\item{3.~~}
$C_{\infty}=Q\cap\{\t=0\}$,

\item{4.~~}
$C_0=\{\x=\eta=0\}\subset Q$.

\noindent The way in which these curves meet each other on $Q$ is indicated in
 figure 1. The points of intersection are:
\item{$\bullet$~~}
$[1,-8,\; 64,\- 0]$, the point of tangency between $C_{\rm con}$ and
$C_\infty$,

\item{$\bullet$~~}
$[1,\- 0,\- 0,\- 0]$, the point of tangency between $C_1$ and $C_\infty$,

\item{$\bullet$~~}
$[0,\- 0,\- 1,\- 0]$, the point of intersection of $C_0$ and $C_\infty$,

\item{$\bullet$~~}
$[0,\- 0,\- 1,\- 1]$, the common point of intersection of $C_0$, $C_1$, and
$C_{\rm con}$, and

\item{$\bullet$~~}
$[1,-4,\; 16,\; 16]$, the intersection point of $C_1$ and $C_{\rm con}$
through which $C_0$ does not pass.
\bigskip
\noindent
The moduli space of the mirror \Mtwo\ has a very similar structure.
The polynomial is
$$
p ~=~ x_1^{12} + x_2^{12} + x_3^6 + x_4^6 + x_5^6
- 12\ps\, x_1x_2x_3x_4x_5
- 2\ph\, x_1^6 x_2^6$$
with the group generated by
$$\eqalign{
&(\IZ_6;~0,5,1,0,0)~,\cr
&(\IZ_6;~0,5,0,1,0)~,\cr
&(\IZ_2;~0,1,0,0,1)~.\cr} $$
The enlarged group includes a $\IZ_{12}$ acting by
$(\ps,\ph)\mapsto (\b\ps,-\ph)$, with $\b$ a nontrivial
twelfth root of unity. The invariants are
$\ps^{12},~\ps^6\ph$ and $\ph^2$. The remaining analysis is
very similar; note that the conifold locus in this case is defined by
$864\, \ps^6 + \ph = \pm 1$.
\subsection{The locus $\ph^2=1$}
Returning to the case of \Mone, we observe that something interesting happens
when we restrict to the locus $\ph=1$. The resulting family of singular
threefolds is birationally equivalent to the mirror family of $\cp5[2,4]$ which
is described in Refs.~
\REFS{\rLT}{A.~Libgober and J.~Teitelbaum, Int.\ Math.\ Res.\ Notices (1993)
29.}
\REFSCON{\rBvS}{V.~Batyrev and D.~van Straten,
``Generalized Hypergeometric Functions and
  Rational Curves on Calabi-Yau Complete Intersections in Toric Varieties'',\\
  alg-geom/9307010.}
\REFSCON{\rperiods}{P.~Berglund, P.~Candelas, X.~de~la~Ossa, A.~Font,
T.~H\"ubsch, D.~Jan\v{c}i\'c and F.~Quevedo, ``Periods for \cy\ and
Landau-Ginzburg Vacua'',
CERN-TH. 6865/93, HUPAPP-93/3, NEIP 93-004, NSF-ITP-93-96, UTTG-13-93,\\
hepth~9308005}
\refsend.
For simplicity we make a rescaling of the
coordinates of the $\cp5[2,4]$ mirror
family given in \cite{\rperiods} so that the equations defining the complete
intersection have the form
$$\eqalign{
y_0^2+y_1^2+y_2^2+y_3^2 &~=~\eta\, y_4y_5\cr
y_4^4 + y_5^4&~=~y_0y_1y_2y_3\cr}\eqlabel{yeqs}$$
where $\eta$ is the parameter.  We must divide this complete intersection by
the group $H$ of coordinate rescalings which preserve both hypersurfaces
as well as the holomorphic 3--form.  Explicitly, this is given by
$$
  \left\{ ~(\a^{4a},\a^{4b},\a^{4c},\a^{4d},\a^e,\a^f)~\Big| ~e{+}f\equiv 0~
(\hbox{mod}\, 8),~~
4a{+}4b{+}4c{+}4d\equiv 4e~(\hbox{mod}\, 8)~\right\}~.$$

To see the birational equivalence, define a rational map
$$
  \Phi : \cp4^{(1,1,2,2,2)}/G~\longrightarrow~\cp{5}/H$$
via
$$\eqalign{y_0 &= x_1^4 - x_2^4\cr
           y_1 &= x_3^2\cr
           y_2 &= x_4^2\cr}\qquad
  \eqalign{y_3 &= x_5^2\cr
           y_4 &= x_1\sqrt{x_3 x_4 x_5}\cr
           y_5 &= \a {x_2 y_4\over x_1}\cr}\eqlabel{xeqs}$$
which is easily seen to be compatible with the actions of $G$ and $H$.
The image $X\subset\cp{5}/H$ of $\Phi$  is defined by the second of equations
\eqref{yeqs}, which is $H$--invariant.  It can then be checked that the
rational map $X\rightarrow\IP^{(1,1,2,2,2)}/G$ given by
$$\eqalign{
            x_1 &= y_4/\sqrt{x_3 x_4 x_5}\cr
            x_2 &= \a^{-1} y_5/\sqrt{x_3 x_4 x_5}\cr
            x_3 &= \sqrt{y_1}\cr
            x_4 &= \sqrt{y_2}\cr
            x_5 &= \sqrt{y_3}\cr}$$
is well-defined, and is a rational inverse of $\Phi$.

  The image of $\{p=0\}/G$ in $X$ is defined by
$$
  y_0^2 + y_1^2 + y_2^2 + y_3^2 - 8\psi\a^{-1} y_4 y_5~=~0$$
which follows from \eqref{xeqs}.  Thus, via $\Phi$, the mirror family with
$\phi = 1$ goes over to the mirror family for the $(2,4)$ complete intersection
after making the substitution $\eta = 8\psi\a^{-1}$.

In the case of \Mtwo , the limit $\ph=1$ analogously leads to
a manifold birationally equivalent to the mirror of
a (2,6) complete intersection in $\IP_5^{(1,1,1,1,1,3)}$.

This is the first hint of a relationship between the quantum geometry of our
two--parameter family, and that of the one--parameter family associated to the
complete intersection. We will later see a relationship between the Yukawa
couplings as well as the geometry of the manifolds themselves.
\subsection{More about the moduli space}
We need a compactification of the moduli space which is smooth,
and whose boundary is a divisor with normal crossings.  This
can be constructed by blowing up the previous compactified moduli
space.  To keep
track of the blowups in an efficient way, it is convenient to
use the language of toric geometry.  (For a review in the physics literature,
see~
\Ref\rMarkushevich{D.~G.~Markushevich, ``Resolution of Singularities (Toric
Method)'', appendix to: D.~G.~Markushevich, M.~A.~Olshanetsky, and
A.~M.~Perelomov, Comm. Math. Phys. {\bf 111} (1987), 247.}).

The singular quadric $Q=\{\xi\zeta-\eta^2=0\}\subset \IP_3$
is isomorphic to the weighted projective space $\IP^{(1,1,2)}$,
and this is the starting point for identifying it as a toric variety.
The toric diagram is the union of the cones whose edges are the rays
spanned by the vectors $(1,0)$, $(0,1)$ and $(-2,-1)$, as illustrated
in figure 2.
For definiteness, we choose coordinates $(u,v)$ associated to the
first quadrant in the toric diagram.  Then the three coordinate
charts are described as follows:  we select the subset of the rational
monomials $u^Av^B$ which satisfy conditions
\item{\ } $A>0$ and $B>0$ in the chart associated to $(1,0)$, $(0,1)$
\item{\ } $A>0$ and $-2A-B>0$ in the chart associated to $(1,0)$, $(-2,-1)$
\item{\ } $-2A-B>0$ and $B>0$ in the chart associated to $(-2,-1)$, $(0,1)$

\noindent
The monomials associated to these are generated by
\item{\ } $u$ and $v$ in the first chart
\item{\ } $v^{-2}u$ and $v^{-1}$ in the second chart
\item{\ } $u^{-1}$, $u^{-1}v$ and $u^{-1}v^2$ (subject to a relation) in the
third chart

\noindent
To make the identification with our moduli space explicit, we choose
$u=\phi^{-2}$, $v=\phi^{-1}\psi^4$.  Then in the third chart (which has
the singularity) we find:
$$u^{-1}=\phi^2=\tilde{\zeta}~,\quad u^{-1}v=\phi\psi^4=\tilde{\eta}~,
\quad u^{-1}v^2=\psi^8=\tilde{\xi}~.$$

In the toric description, each vector $\vec{v}$ in the toric diagram is
associated to a divisor $D_{\vec{v}}$ on the toric variety.  The divisor
$D_{(0,1)}$ is described by $v=0$, and so coincides with the curve $C_0$.
The divisor $D_{(1,0)}$ is described by $u=0$, and corresponds to the
compactification divisor $C_\infty$ from before.  The divisor $D_{(-2,-1)}$
is the locus $\phi=0$, which plays no particular role in our discussion
of the moduli space.

The curve $C_1=\{\phi^2=1\}$ is given by $u=1$ in the first chart,
and by $uv^{-2}=(v^{-1})^2$ in the second chart.  Its point of tangency
with $C_\infty$ is therefore located at the origin in the second chart.

We can now do several toric blowups.  First, to resolve the singularity
of $Q$ itself, we blow up the singular point, which adds the vector
$(-1,0)$ to the toric diagram.  We then blowup twice in the second
chart (to replace the tangency between $C_1$ and $C_\infty$ by normal
crossings divisors),
adding the vectors $(-1,-1)$ and then $(0,-1)$ to the toric
diagram.  (The resulting toric diagram after all blowups is shown in
figure 3.)
To simplify notation, we will use the same symbol to denote a
curve as well as its proper transform via the blowup map.
The resulting surface contains a ``chain'' of divisors
$D_{(-1,0)}$, $D_{(0,1)}=C_0$, $D_{(1,0)}=C_\infty$,
$D_{(0,-1)}$, $D_{(-1,-1)}$
(with each divisor meeting
the next one in the chain) whose self-intersections are
$-2$, $0$, $0$, $-1$, and $-2$, respectively.

We still must blowup the common point of intersection of $C_0$, $C_1$,
and $C_{\rm con}$, and blowup (twice) the
point of tangency between $C_{\rm con}$
and $C_\infty$; these are not toric blowups.  We label the exceptional
curves for these blowups $E_0$, $E_1$, and $E_2$, respectively.
The resulting diagram of curves is depicted in figure 4.

The moduli space of the mirror of $\M_2=\Mtwo$ has a very similar structure.
\newpage
\section{new}{Monodromy and the Large Complex Structure Limit}
\vskip-20pt
\subsection{The large complex structure limit}
The first step in determining the mirror map is to locate the point
corresponding to the large complex structure limit.
To this end we wish to make some observations about the nature
of the large complex structure limit (cf.~
\REFS{\rCDFLL}{A.~Ceresole, R.~D'Auria, S.~Ferrara, W.~Lerche and
       J.~Louis,  Int. J. Mod. Phys. {\bf A8} (1993) 79.}
\REFSCON{\rcompact}{D.~R.~Morrison, ``Compactifications of Moduli Spaces
Inspired by Mirror Symmetry'', DUK-M-93-06.}
\refsend).
Consider the general case of $n$
parameters $t^j~,~j=1,\ldots,n$. These may be chosen such that, in the
\lcsl\ the prepotential assumes the form
$$
\ca{F}=-{1\over 3!}\yzero_{ijk}t^i t^j t^k - {1\over 2}q_{ij} t^i t^j -
l_i t^i - \x+ \cdots~, \qquad\boxed{$w^0=1$}$$
where the $\yzero_{ijk}$ are the topological values for the Yukawa
couplings
and the ellipsis denotes terms that are exponentially small and are also
periodic under $t^i\to t^i+1$. We have also chosen the gauge $w^0=1$.

We form the period vector
$$
\ip =\pmatrix{\ca{F}_0\cr
              \ca{F}_i\cr
                     1\cr
                   t^j\cr}~,$$
where $\ca{F}_i=\pd{\ca{F}}{t^i}$ and $\ca{F}_0=2\ca{F}-t^i\ca{F}_i$, and
we
consider the effect of making the replacement $t^j\to t^j+\d^j_i$ on
$\ip$.
This corresponds to a matrix $S_i$
$$
\ip\to S_i\ip ~~~~ \hbox{with}~~~~
S_i=\pmatrix{
1&   -\d^T_i& {1\over 6}\yzero_{iii}-2l_i&\half\yzero^T_{ii}+q^T_i\cropen{3pt}
0&\- {\bf 1}& -\half \yzero_{ii}+q_i     &            -\yzero_i\cropen{3pt}
0&      \- 0&                           1&                 \- 0\cropen{3pt}
0&      \- 0&                        \d_i&           \- {\bf 1}\cropen{3pt}}
{}~.$$
Here $\d_i$, $q_i$ and $\yzero_{ii}$ are vectors and the $\yzero_i$ are
matrices
$$\eqalign{
\d_i=(\d^j_i)~~~,~~~q_i&=(q_{ij})~~~,~~~\yzero_{ii}=(\yzero_{iij})\cr
\yzero_i&=(\yzero_{ijk})~.\cr}$$
Note that the $S_i$ are integral matrices for suitable $q_{ij}$ and $l_i$,
this
being possible in virtue of the fact that the intersection numbers
$\yzero_{ijk}$
for $H^2(\M,\IZ)$ satisfy
$$
\yzero_{iij} = \yzero_{ijj}~({\rm mod}~2).$$
We could in fact have improved the appearance of the $S_i$ by departing
from
the usual convention and writing the $w$'s in the order $(w^j,w^0)$ in
$\ip$
then the $S_i$ would be upper triangular.
Set now
$$
R_i=S_i - {\bf 1}$$
and observe that
$$
\llap{\hbox{$\eqalign{i.&\cr ii.&\cr iii.&\cr}$\hskip50pt}}
\eqalign{\left[ R_i,R_j \right] &= 0\cr
R_iR_jR_k &= \yzero_{ijk}Y\cr
R_iR_jR_kR_l &= 0\cr}\eqlabel{lcsl}$$
where $Y$ is a matrix independent of $i$. (In this basis $Y$ is the matrix
which has $Y_{1,n+1} = 1$ and all other components
zero.)
The utility of \eqref{lcsl} is that these relations give a
characterisation of the \lcsl\ independent of the choice of basis for
the periods. Under a change of basis the only thing that changes is
the form of the matrix $Y$.

We anticipate that the \lcsl\ consists, in the general case, of $n$
codimension 1 hypersurfaces in the (compactification of the) moduli space
meeting transversely in a point and such that the monodromies of the
period
vector about these hypersurfaces correspond to the properties
\eqref{lcsl}.
There will thus be periods such that $S_i$ corresponds to
$t^j\to t^j + \d^j_i$.
One of us has speculated elsewhere~
\cite{\rcompact}\
that the $t^j$ may in fact correspond to an integer basis. However, we
will not assume this.

We first attempt to locate possible large complex structure limits
by using \eqref{lcsl}.  We need to consider points in a compactification
of the moduli space which are at the transverse intersection of $n$
boundary divisors (when the number of parameters is $n$), and calculate
monodromy matrices $S_i$ around each of those boundary divisors.
For technical reasons having to do with the known structure of
monodromy matrices~
\Ref{\rLandman}{A.~Landman, Trans. Amer.  Math. Soc. {\bf 181} (1973), 89.},
we must be willing to replace $S_i$ by a positive integral power
$(S_i)^{k_i}$.  Then $R_i=(S_i)^{k_i}-{\bf 1}$ is a nilpotent matrix, and we
can define a logarithm
$$N_i=\log (S_i)^{k_i} = R_i - {1\over2}R_i^2 + {1\over3}R_i^3 - \cdots$$
by a finite power series.
A key property we will use to identify large complex structure limit points
is that the $R_i$ are linearly independent (to ``mirror'' the
corresponding property of a set of generators for the K\"ahler cone)
and that for a general linear combination
$$R=a_1R_1+\ldots +a_nR_n,$$
we have $R^3\neq 0$.  Equivalently, we can check this using the logarithms:
for a general combination
$$N=a_1N_1+\cdots+a_nN_n$$
we have $N^3\ne0$.
We also anticipate by mirror symmetry that the linear transformations
corresponding to the matrix $N_i$ may be identified (in the large radius
limit) with the linear
tranformation on $\sum_{i=0}^3H^{2i}(\ca{M})$ given by cup product with the
corresponding class $\o_i\in H^2(\ca{M})$.
\subsection{Monodromy calculations}
In this subsection,
we use the differential equations satisfied by the cohomology
classes of $\ca{M}$ to calculate monodromy
around the  divisors of the compactification of the moduli space described in
Section~\chapref{moduli1}.  Only the case of $\wP[8]$ will be considered here,
but the discussion can be immediately adapted to $\WP[12]$.

These differential equations can be obtained as explained in
Refs.
\REF{\rCF}{A.~C.~Cadavid and S.~Ferrara, \plb{267} (1991) 193.}
\REF{\rBV}{B.~Blok and A.~Varchenko, Int. J. Mod. Phys. A{\bf 7},
      (1992) 1467.}
\REF{\rLSW}{W.~Lerche, D.~J.~Smit and N.~P.~Warner, \npb{372} (1992)
       87.}
\cite{{\rMorrison,\rFont,\rCDFLL,\rCF--\rLSW}}.
In the notation of \cite{\rFont},
we choose the basis for $H^3(\ca{M})$ corresponding
to the choice of monomials
$$\displaylines{
x_0x_1x_2x_3x_4x_5\crm
x_0^2x_1^2x_2^2x_3^2x_4^2x_5^2 ~~~~~~ x_0^2x_1^5x_2^5x_3x_4x_5 \crm
x_0^3x_1^3x_2^3x_3^3x_4^3x_5^3 ~~~~~~ x_0^3x_1^6x_2^6x_3^2x_4^2x_5^2 \crm
x_0^4x_1^7x_2^7x_3^3x_4^3x_5^3\cr}$$

The differential equations take the matrix form
$$\pd{R}\psi = RM_{\psi}~,~~~~\pd{R}\phi = RM_{\phi}
\eqlabel{mms}~. $$
Said differently, the matrix of differential forms
$-(M_{\psi}\,d\psi+M_{\phi}\,d\phi)$ is the
connection matrix for the local
system defined by the varying cohomology spaces in the chosen basis.
This local system is only defined over a cover of the moduli space
described by $(\psi,\phi)$ coordinates.

The matrices $M_{\psi}, M_{\phi}$ can be determined as described
in Refs. \cite{{\rFont,\rCF}}. Notice that
they must satisfy the integrability condition
$$
\left[ M_{\phi}, M_{\psi} \right] =
\pd{M_{\phi}}\psi -
\pd{M_{\psi}}\phi$$
We find
$$
M_{\psi} = \pmatrix{
\- 0 &\-  0 &\-  0 &\-  0 &\-  0 &\-  \ds{ {{\psi}\over {64\D}} } \crm
  -8 &\-  0 &\-  0 &\-  0 &\-  0 &\ds{- {{15\psi^2}\over {8\D}} } \crm
\- 0 &\-  0 &\-  0 & \ds{ -\psi } &\-  0 &
\ds{- {{\psi(44\psi^4 +3\phi)}\over {4\D}} } \crm
\- 0 &   -8 &\-  0 &\-  0 &\-  0 &\-  \ds{ {{25\psi^3}\over {\D}} } \crm
\- 0 &\-  0 &   -8 &\-  \ds{ 24\psi^2 } &\-  0 &
\- \ds{ {{2\psi^2(148\psi^4 + 13\phi)}\over {\D}} } \crm
\- 0 & \- 0 &\-  0 & \ds{ -64\psi^3 } & -8 &
\ds{- {{128\psi^3(8\psi^4 +\phi)}\over {\D}} }
}$$

\bigskip
$$
M_{\ph} = \pmatrix{
\- 0 &\-  0 & -\ds{1\over 32Z} &\-  0 &\-  0
&\-  \ds{{\ps^2(4\ps^4 + \ph)\over 128 Z\D} } \crm
\- 0 &\-  0 &\-  \ds{3\ps\over 4Z} &\-  0 & -\ds{1\over 8Z}
& \ds{-{15\ps^3(4\ps^4 + \ph)\over 16 Z\D} } \crm
-2 &\-  0 &\-  \ds{3\ph\over 2Z} &\-  0 & -\ds{\ps^3\over 4Z}
& \ds{-{\a\ps^2\over 8 Z\D} } \crm
\- 0 &\-  0 & -\ds{2\ps^2\over Z} &\-  0 &\-  \ds{5\ps\over 4Z}
& \- \ds{{\b\over 32 Z\D} } \crm
\- 0 & -2 & -\ds{4\ps\ph\over Z} &\-  0 &\-  \ds{2(3\ps^4+\ph)\over Z}
&\-  \ds{{\g\ps^3\over  Z\D} } \crm
\- 0 &\-  0 &\-  0 & -2 & -\ds{4\ps(4\ps^4+\ph)\over Z}
& \ds{-{\d\over 2 Z\D} }
\crs}$$
where we have defined
$$\eqalign{
\D &= (8\ps^4+\ph)^2 -1 \crm
 Z &= 1 - \ph^2 \crm
\a &= 3(4\ps^4 +\ph)^2 + 2Z \cr}~~~~~~~~
\eqalign{
\b &= 25Z + 16 \D \crm
\g &= 7Z + (4\ps^4 + \ph)(36\ps^4 + 13\ph) \crm
\d &= 8Z(8\ps^4+\ph) + 3\D(4\ps^4 + \ph)\cr}$$

However, our chosen
basis is not invariant under the group of phase symmetries.  To achieve
invariance and hence describe a local system over the true moduli space,
the basis must be multiplied by the respective quantities
$$
\psi,\ \psi^2,\ \psi^5,\ \psi^3,\ \psi^6,\ \psi^7.$$
This is achieved by using the Leibniz rule:
$$
d(\psi^r e_j)=\psi^r\,de_j+r\psi^{r-1} e_j\,d\psi,$$
where $\{e_j\}_{j=1,\ldots,6}$ is the original basis in which
$M_{\ps}$ and $M_{\ph}$ were computed.  After effecting
this change of basis, the connection matrix now becomes single valued in
the coordinates $\tilde{\x},\ \tilde{\eta},\ \tilde{\z}$ defined in
Section~\chapref{moduli1}.

The monodromy can be found by the methods of Deligne~
\REFS{\rDeligne}{P.~Deligne, {\it Equations Diff\'erentielles \`a Points
Singuliers R\'eguliers}, Lecture Notes in Mathematics {\bf 163},
Springer-Verlag
1970.}
\refsend.
Since our basis is not flat, the monodromy we calculate is interpreted
as an automorphism of the local system, rather than an automorphism of
a fixed vector space.
We use the blown up model of the moduli space constructed in Section 3.3.
This is a smooth space, and the boundary is a normal crossings divisor.
If we
 restrict attention to a situation locally near the intersection $p$
of two normal crossings boundary divisors $D_1,\ D_2$, and choose coordinates
$(s_1,s_2)$ near $p$ such that each $D_i$ is locally defined by $s_i=0$, then
we can find the monodromy about $D_1$ by setting $s_2=c$, where $c$ is
a non-zero constant.
So we are reduced to calculations involving local systems on the punctured
disk.

According to the ``regularity theorem''~
\Ref{\rregularity}{N. Katz, Publ.\ Math.\ I.H.E.S., {\bf 39} (1971) 175},
the differential equation \eqref{mms} has regular singular points.
It is shown in \cite{\rDeligne}
that for such an equation,
a basis can be found for which the connection matrix
has at worst first order poles (and an algorithm is given there if needed).
Its matrix of residues is easily calculated.  The eigenvalues of the residue
matrix are necessarily rational,
and will in fact be integers if and only if the monodromy is unipotent.
Supposing then that the monodromy is in fact unipotent, the eigenvalues will
be integers.  A basis can be chosen for which the eigenvalues will all be
zero (the resulting extension of the local system to the puncture is called the
{\sl canonical extension\/} in \cite{\rDeligne}).
If $\rm{Res}$ denotes the residue matrix, then the monodromy transformation
will be conjugate to
$S=e^{-2\pi i{\rm Res}}$.  So the index of unipotency of $S$
can be read off from the Jordan form of $\rm{Res}$.
This has been done near each of
the normal crossing boundary points of the
blowup of the moduli space described in Section~\chapref{moduli1}.

Let $D_{\vec{v}}$ denote the toric divisor associated to an edge spanned by
the vector $\vec{v}$. The results of the calculations may be summarized as
follows.
\medskip
$$\vbox{\offinterlineskip\halign{
\strut # height 15pt depth 8pt& \quad$#$\quad\hfill\vrule&
\quad$#$\hfill&\quad # \quad\hfill\vrule\cr
\noalign{\hrule}
\vrule & \hbox{Curve} &&\hskip10pt\hbox{Monodromy}\cr
\noalign{\hrule}
\vrule &D_{(1,0)}   & S   & unipotent of index 2   \cr
\vrule &D_{(0,-1)}  & S   & unipotent of index 4   \cr
\vrule &D_{(-1,-1)} & S^2 & unipotent of index 4   \cr
\vrule &D_{(-1,0)}  & S^8 & \hskip-7pt $={\bf 1}$       \cr
\vrule &D_{(0,1)}   & S^4 & unipotent of index 2   \cr
\vrule &C_{\rm con} & S   & unipotent of index 2   \cr
\vrule &C_1         & S^2 & unipotent of index 2   \cr
\vrule &E_0         & S^4 & unipotent of index 2   \cr
\vrule &E_1         & S   & unipotent of index 2   \cr
\vrule &E_2         & S   & unipotent of index 2   \cr
\noalign{\hrule}
}}$$
\medskip
We sketch the calculation for one of the boundary divisors.  The other
calculations are similar (and for the most part easier).  We compute
the Jordan form of the monodromy about $C_{\rm con}$ near the point
where $\ph=0$.  Here, we have local coordinates
$s_1=(8\ps^4+\ph)^2-1$ and
$s_2=\ph^2$, with $s_1=0$ the local equation of
$C_{\rm con}$.
To get the monodromy about $C_{\rm con}$, we need the coefficient of
$ds_1/s_1$ in the connection matrix, after changing to $(s_1,s_2)$
coordinates (that is, we calculate the residue along $C_{\rm con}$).

The result is
$$\pmatrix{
0 & 0 & 0 & 0 & 0 &\-\ds{\ps^4\over 4096(8\ps^4+\ph)}  \crm
0 & 0 & 0 & 0 & 0 &  \ds -{15\ps^4\over 512(8\ps^4+\ph)}\crm
0 & 0 & 0 & 0 & 0 &  \ds -{44\ps^4+3\ph\over 256(8\ps^4+\ph)}\crm
0 & 0 & 0 & 0 & 0 &\-\ds {25\ps^4\over 64(8\ps^4+\ph)} \crm
0 & 0 & 0 & 0 & 0 &\-\ds {148\ps^4+13\ph\over 32(8\ps^4+\ph)} \crb
0 & 0 & 0 & 0 & 0 &  \ds -2 \cr}~.$$
The eigenvalues are 0 and $-2$, hence the basis does not extend to a basis
of the canonical extension, and this cannot be the logarithm of the monodromy.
But this is easily remedied by changing the last basis element by multiplying
it by $s_1^2$ (to remove the eigenvalue of $-2$), and adding on an appropriate
linear combination of the other basis elements so that the residue
of the connection matrix in this basis becomes strictly upper triangular.
This is a straightforward
calculation. After doing this, the resulting matrix has Jordan form
$$\pmatrix{
0 & 1 & 0 & 0 & 0 & 0 \cr
0 & 0 & 0 & 0 & 0 & 0 \cr
0 & 0 & 0 & 0 & 0 & 0 \cr
0 & 0 & 0 & 0 & 0 & 0 \cr
0 & 0 & 0 & 0 & 0 & 0 \cr
0 & 0 & 0 & 0 & 0 & 0 \cr
},\eqlabel{jordan}$$
hence the corresponding monodromy transformation is unipotent of index 2.

We now apply the criterion stated at  the end of Section 4.1.
In our situation,
we observe that among the divisors, only
$D_{(0,-1)}$ and $D_{(-1,-1)}$ have monodromy transformations with a power that
is  maximally unipotent.  The others all have index 2.
$D_{(0,-1)}$ only meets the divisors
$D_{(1,0)},\ C_1,\ \hbox{and }D_{(-1,-1)}$, and $D_{(-1,-1)}$ only meets
the divisor $D_{(1,0)}$.  All other intersections are intersections
of divisors with index 2 monodromy.  At a point $p$
of intersection
of two divisors with unipotent monodromy of index 2, let $N_1$ and $N_2$ be
the respective monodromy logarithms.  Since $(a_1N_1+a_2N_2)^3=0$ for all
$a_1$ and
$a_2$, the point $p$ cannot be a large complex structure limit point.  This
leaves only 3 candidates for the large complex structure limit, the three
points where other divisors $D_{(1,0)},\ C_1,\ \hbox{and }D_{(-1,-1)}$
meet $D_{(0,-1)}$.  But the last two can be ruled out.  To rule out
$D_{(-1,-1)}$ it is easy to see from the local geometry of the moduli space
that the monodromy about $D_{(0,-1)}$ is the square of the monodromy about
$D_{(-1,-1)}$
(cf.~
\REFS{\rBK}{P.~Berglund and S.~Katz, In preparation}
\refsend).  So the monodromy logarithms are proportional, hence cannot yield
a \lcsl\ point.  To rule out the intersection with $C_1$,
it suffices to observe that
our monodromy calculation shows
that the Jordan decomposition of the logarithm of the
square of the monodromy transformation about $C_1$ is also given by
\eqref{jordan}.
If the corresponding intersection were a \lcsl, then the corresponding $(1,1)$
form on the mirror would annihilate a 5~dimensional subspace of
$\sum_{i=0}^3H^{2i}(\ca{M})$, which would then have to be
$\sum_{i=1}^3H^{2i}(\ca{M})$ by degree considerations and the fact that
the form cannot annihilate the generator of $H^0(\ca{M})$.  But this is
impossible for a K\"ahler manifold.

So there is only one \lcsl\ point, the intersection of $D_{(1,0)}$ and
$D_{(0,-1)}$.
\newpage
\section{more}{Considerations of Toric Geometry.}
There is another strategy for locating the large complex structure
points:  the {\sl monomial-divisor mirror map} introduced in~
\REF{\rMDMM}{P.~S. Aspinwall et al.,
``The Monomial-Divisor Mirror Map'', IASSNS-HEP-93/43, in preparation.}
\cite{{\rAGM,\rMDMM}}.
  (To see more details of the theory in a general situation,
the reader is referred to those papers.)
This map is  a refinement of  Batyrev's proposal~
\Ref{\rBatyrev}{V.~V.~Batyrev, ``Dual polyhedra and mirror symmetry for \cy\
hypersurfaces in toric varieties'', preprint, November 18, 1992.}\
for mirror symmetry for toric hypersurfaces, which generalizes the
toric interpretation~
\Ref{\rRoan}{S.-S. Roan, Internat.\ J.\ Math.\
  {\bf 2} (1991) 439.}\
of the orbifolding construction of mirror symmetry \cite{\rGP}.

We start by considering the desingularization of \wP\
as a toric variety.  Embed the torus $T=(\IC^*)^4$ in \wP\ via
$$
(\t_1,\t_2,\t_3,\t_4)\mapsto (1,\t_1,\t_2,\t_3,\t_4)~.$$
As usual, let $N=\Hom(\IC ^*,T)$ be the lattice of 1~parameter subgroups of
$T$, and let $M=\Hom(T,\IC ^*)$ be the dual lattice of characters of $T$.
Put $M_\IR =M\otimes\IR $, and $N_\IR =N\otimes\IR $.
This variety has
its $N_\IR $ fan generated by the edges spanned by the vectors
$$\matrix{(  -1,&  -2,&  -2,&  -2)\cr
          (\- 1,&\- 0,&\- 0,&\- 0)\cr
          (\- 0,&\- 1,&\- 0,&\- 0)\cr
          (\- 0,&\- 0,&\- 1,&\- 0)\cr
          (\- 0,&\- 0,&\- 0,&\- 1)\cr
          (\- 0,&  -1,&  -1,&  -1)\cr}$$
where each row contains the coordinates of an edge.
The set of these edges will be denoted by $S$.
Ordering has been
chosen so that the first 5 edges, in order, correspond to the proper transforms
of $x_i=0$ under the identification of edges in the fan with divisors
in the toric variety which are invariant under the torus.  The edge spanned
by the last vector
corresponds to the exceptional divisor.  Note that its coordinates are the
average of the coordinates of the first two edges; this corresponds to the
fact that $x_1=x_2=0$ has been blown up.

Let $\IZ ^S$ denote the free abelian group on $S$, which may be thought of as
the group of $\IZ$-valued functions on $S$.  This may also be thought of as
the group of $T$-invariant divisors on \wP.
Linear equivalence is
identified with the sublattice $M$ of $\IZ ^S$ induced by
viewing elements of $M$ as $\IZ $-linear functionals on
$N$.  So the space of linear equivalence classes of
divisors
is naturally the quotient lattice $P=\IZ ^S/M$.  By duality, the dual
lattice $P^*$ is the natural space for coordinate functions on the
vector space $P$.  This is
naturally the sublattice of $(\IZ ^S)^*$ which is annihilated by the transpose
of the inclusion of  $M$ in $\IZ ^S$.  We choose the $\IZ $-basis for
$P^*$ spanned by the two vectors in the rows of
$$
\pmatrix{
1 &\- 1 &\- 0 &\- 0 &\- 0 &  -2 \cr
0 &\- 0 &  -1 &  -1 &  -1 &  -1 \cr}
$$
We choose these two vectors as defining coordinate functions on $P$.
The coordinates of the 6 divisor classes
can be read off as the 6 columns of the
above matrix. Note also that the divisor class $L$ can be represented by
$x_1 = 0$ and so corresponds to the vector $(1,0)$ while the divisor class $H$
can be represented by $x_3 = 0$ and corresponds to $(0,-1)$.  Thus, the fourth
quadrant is identified with the \K-cone.  See also \cite{{\rAGM,\rMDMM}}.

We now consider the moduli space of the mirror family of $\Mone$.  The most
general $G$--invariant hypersurface of degree $8$ on $\cp4^{(1,1,2,2,2)}$ has
an
equation of the form
$$
c_1x_1^8+c_2x_2^8+c_3x_3^4+c_4x_4^4+c_5x_5^4+c_6x_1x_2x_3x_4x_5+
c_7x_1^4x_2^4=0. \eqlabel{family}
$$

In our context, we can interpret the monomial-divisor mirror map as
 saying that there is a natural $T$ action on the 4~dimensional
$G$-quotient space which contains the mirror hypersurfaces such that
the weights of the seven monomials appearing in equation \eqref{family}\ are
exactly $S\cup\{(0,0,0,0)\}$.
The reader can check that this action is given by
$$(\t_1,\t_2,\t_3,\t_4)\cdot(x_1,x_2,x_3,x_4,x_5)
=((\t_1\t_2^2\t_3^2\t_4^2)^{-1/8} x_1,\,
\t_1^{1/8}x_2,\t_2^{1/4}x_3,\, \t_3^{1/4}x_4,\, \t_4^{1/4}x_5)~.$$
The fractional exponents arise only because we have we have related everything
to the symbols $x_i$, which are not $G$-invariant.
The isomorphism type of this weighted hypersurface is left unchanged by
rescaling of the coordinates via $T$, together with an overall rescaling
of all the $c_i$ by a constant.  Let $\barT=T\times\IC ^*$, a 5~dimensional
torus.  We can incorporate the rescaling of the $c_i$ by appending a 1
to the list of $T$-weights, obtaining the \barT\ weights
$$
\pmatrix{
  -1  &   -2  &   -2  &   -2  &\-  1\cr
\- 1  &\-  0  &\-  0  &\-  0  &\-  1\cr
\- 0  &\-  1  &\-  0  &\-  0  &\-  1\cr
\- 0  &\-  0  &\-  1  &\-  0  &\-  1\cr
\- 0  &\-  0  &\-  0  &\-  1  &\-  1\cr
\- 0  &   -1  &   -1  &   -1  &\-  1\cr
\- 0  &\-  0  &\-  0  &\-  0  &\-  1\cr}\eqlabel{edges}$$

Using bars to denote a toric construction for $\barT$, and letting
\barS\ denote the set of 7~vectors in \barN\ from \eqref{edges}, we
see that
the moduli space is a quotient of $\IC ^{\barS}$ by $(\IC ^*)^5$.
This contains the 2~torus $T'=(\IC ^*)^{\barS}/(\IC ^*)^5$; we will
use primes for toric constructions for $T'$.  So the moduli
space is a $T'$-toric variety.  Proceeding similarly to the discussion for
$P$, we can construct coordinates from the vectors
$$
\pmatrix{
\- 1 &\-  1 &\-  0 &\-  0 &\-  0 &   -2  &\-  0\cr
\- 0 &\-  0 &   -1 &   -1 &   -1 &   -1  &\-  4\cr}
$$
and the fan of our toric moduli space is spanned by the edges
$$(1,0),~(0,-1),~(-2,-1),~(0,4)~.$$
We may as well use $(0,1)$ in place of $(0,4)$, as it spans the same
edge.
(The toric diagram is shown in figure 5.)
This is a special case of the secondary
fan discussed by Oda and Park~
\Ref\rOP{T. Oda and H.~S. Park, T\^ohoku Math. J. {\bf 43} (1991) 375};
 and this is a general situation
arising in mirror symmetry for toric hypersurfaces
as described in \cite{{\rAGM,\rMDMM}}.
By comparison with our discussion of the K\"ahler moduli space, the large
complex structure limit corresponds to the cone spanned by $(1,0),~(0,-1)$,
that is, to the point of intersection of $D_{(1,0)}=C_{\infty}$ with
$D_{(0,-1)}$.

To calculate in coordinates, let $(u,v)$ be coordinates on the affine
part of the toric variety corresponding to $(1,0),(0,1)$.  In general,
the coordinate functions on an affine toric variety are given by the
elements of $M$ which pair non-negatively with the elements of its cone
in $N$, under the natural pairing of $M$ and $N$.  Note that $u$ and $v$
correspond to the elements $(1,0)$ and $(0,1)$ of $M$.  Carrying out
this calculation, we get a collection of local coordinates
in the four regions of the moduli space.  This appears in
Table~\tabref{fourregs1}.
\def\shortrule{\vrule height 12pt depth 8pt}
$$\vbox{\offinterlineskip\halign{
#& \hskip20pt$#$\hskip20pt\hfil\vrule
& \hskip20pt$#$\hskip20pt\hfil\vrule
& \hskip20pt$#$\hskip20pt\hfil\vrule\cr
\noalign{\hrule}
\shortrule & \hbox{Edges of Cone} & \hbox{$(u,v)$ Coordinates}
           & \hbox{$(\psi,\phi)$ Coordinates}\cr
\noalign{\hrule\vskip3pt\hrule}
\shortrule & $$(1,0)~,~(0,-1)$$ & $$(u,v^{-1})$$ &
   $$(\phi^{-2},\phi\psi^{-4})$$\cr
\noalign{\hrule}
\shortrule & $$(0,-1)~,~(-2,-1)$$ & $$(u^{-1},v^{-1},uv^{-2})$$
&$$(\phi^2,\phi\psi^{-4},\psi^{-8})$$\cr
\noalign{\hrule}
\shortrule & $$(-2,-1)~,~(0,1)$$ & $$(u^{-1},u^{-1}v,u^{-1}v^{2})$$ &
        $$(\phi^2,\phi\psi^4,\psi^8)$$\cr
\noalign{\hrule}
\shortrule & $$(0,1)~,~(1,0)$$ & $$(u,v)$$ &
$$(\phi^{-2},\phi^{-1}\psi^4)$$\cr
\noalign{\hrule}
}}
$$
\nobreak
\tablecaption{fourregs1}{The local coordinates for the four regions of the
mirror moduli space for $\Mone$.}
\vskip20pt
\noindent The regions listed second and third are singular; note the relations
$$(u^{-1})(uv^{-2})=(v^{-1})^2$$
and
$$(u^{-1})(u^{-1}v^{2})=(u^{-1}v)^2$$
in the respective regions.

Comparing the coordinates in the third region with the multi-valued
coordinates $\phi$ and $\psi$, we see that the coordinates $u$ and $v$
are expressed in terms of these as \hbox{$u=\phi^{-2}$},
\hbox{$v=\phi^{-1}\psi^{4}$}.
In terms of $\phi$ and $\psi$, the coordinates in the first region are
$\phi^2,\phi\psi^4,\psi^8$, as found earlier.
Two of our charts here already appear in the $\IP^{(1,1,2)}$ model
of the moduli space.

The calculation for $\WP[12]$ is similar.  The only difference worth mentioning
is that now we have $u=\phi^{-2}$ and $v=\phi^{-1}\psi^{6}$.
The results are summarized in Table~\tabref{fourregs2}.

$$\vbox{\offinterlineskip\halign{
#& \hskip20pt$#$\hskip20pt\hfil\vrule
& \hskip20pt$#$\hskip20pt\hfil\vrule
& \hskip20pt$#$\hskip20pt\hfil\vrule\cr
\noalign{\hrule}
\shortrule & \hbox{Edges of Cone} & \hbox{$(u,v)$ Coordinates}
           & \hbox{$(\psi,\phi)$ Coordinates}\cr
\noalign{\hrule\vskip3pt\hrule}
\shortrule & $$ (1,0),(0,-1)$$  & $$ (u,v^{-1})$$
           & $$ (\phi^{-2},\phi\psi^{-6})$$ \cr
\noalign{\hrule}
\shortrule & $$ (0,-1),(-2,-1)$$  & $$ (u^{-1},v^{-1},uv^{-2})$$
           &$$ (\phi^2,\phi\psi^{-6},\psi^{-12})$$ \cr
\noalign{\hrule}
\shortrule & $$ (-2,-1),(0,1)$$  & $$ (u^{-1},u^{-1}v,u^{-1}v^{2})$$
           & $$ (\phi^2,\phi\psi^6,\psi^{12})$$ \cr
\noalign{\hrule}
\shortrule & $$ (0,1),(1,0)$$  & $$ (u,v)$$
           & $$ (\phi^{-2},\phi^{-1}\psi^6)$$ \cr
\noalign{\hrule}
}}
$$
\nobreak
\tablecaption{fourregs2}{The local coordinates for the four regions of the
mirror moduli space for \Mtwo.}
\newpage
\section{fundper}{The Periods}
\vskip-20pt
\subsection{The fundamental period}
An important observation in the present work as indeed in~
\REF{\ralgebraictori}{V.~V.~Batyrev, Duke Math.\ J.\ {\bf 69}  (1992) 349.}
\cite{{\rCdGP,\rperiods,\ralgebraictori}}\
is that it is possible to find the periods of the holomorphic three--form by
direct integration. Let $B_0$ be the cycle
$$\eqalign{
B_0 =
\Big\{ x_k~\big|~x_5&=\hbox{const.}~,~|x_1|=|x_2|=|x_3|=\d~,\cropen{-3pt}
&\hbox{$x_4$ given by the solution to $p(x)=0$ that tends to zero as
$\ps\to\infty$.}~\Big\}\cr}$$
and take for the holomophic three--form the quantity
$$
\O = - {\vert G \vert (\ps\, d) \over (2\p i)^3}\,
{x_5 dx_1dx_2dx_3 \over \pd{p}{x_4}}~.$$
Certain factors have been introduced to simplify later expressions. The
factor $\vert G \vert$ is the order of the group $G$, $d$ denotes the degree of
$p$ and
the factor of $\ps$ ensures that $\O$ is
invariant under
the extended group $\widehat{G}$ and that the fundamental
period, which we shall now define, tends to unity as $\ps\to\infty$.

We define the fundamental period to be
$$\eqalign{
\vp_0(\ph,\ps)
&= \hphantom{-{\ps\,d\over (2\p i)^4}}\int_{B_0}\O\cropen{10pt}
&= -{\ps\,d\over (2\p i)^4}\int_{\g_1\times\cdots\times\g_4}
{x_5dx_1dx_2dx_3dx_4\over p}\cropen{10pt}
&=-{\ps\,d\over (2\p i)^5}\int_{\g_1\times\cdots\times\g_5}
{dx_1dx_2dx_3dx_4dx_5\over p}\cr}$$
where $\g_j$ denotes the circle $|x_j|=\d$. In passing to the second
equality a
factor equal to $\vert G \vert$ is absorbed owing
to the identifications. We also use the fact that
$\left(\pd{p}{x_4}\right)^{-1} =
{1\over 2\p i}\int_{\g_4} {dx_4\over p}$
in virtue of the definition of $B_0$ which
is such that the value of $x_4$ for which $p$ vanishes
tends to zero as $\ps\to\infty$ and hence lies inside
$\g_4$ for sufficiently large $\ps$.
The third equality follows on account of the fact
that the second integral is, despite appearances, independent of $x_5$ in
virtue of the scaling properties of the integrand. We may therefore
introduce
unity in the guise of ${1\over 2\p i}\int_{\g_5}{dx_5\over x_5}$.

In order to perform the integration we consider the case of \Mone \
in which
$$
p_0 = x_1^8 + x_2^8 + x_3^4 + x_4^4 + x_5^4$$
and expand the quantity $1\over p$ in inverse powers of $\ps$
$$
\vp_0= {1\over (2\p i)^5}\int_{\g_1\times\cdots\times\g_5}
{dx_1dx_2dx_3dx_4dx_5 \over x_1x_2x_3x_4x_5}
\sum_{m=0}^\infty{(p_0 -2\ph x_1^4 x_2^4)^m
\over (8\ps)^m (x_1x_2x_3x_4x_5)^m}~.$$
The only terms that contribute in this expression are the terms in the
expansion of \hbox{$(p_0 -2\ph x_1^4 x_2^4)^m$} that cancel the factor of
$(x_1x_2x_3x_4x_5)^m$ in the denominator. A little thought now reveals the
fundamental period to be given by the expression
$$
\vp_0(\ph,\ps) = \sum_{r,s = 0}^\infty
{(8r+4s)!\, (-2\ph)^s \over \left( (2r+s)!\right)^3 (r!)^2\, s!\,
(8\ps)^{8r+4s}} \eqlabel{pi0double}$$
which converges for sufficiently large $\ps$.

Before continuing, we would like to get a closer look at the periods for
$\phi=0$.  In this case we have
$$\eqalign{\varpi_0(\psi,0)
&= \sum_{n=0}^{\infty} {(8n)!\over {((2n)!)^{\,3} \, n!^{\,2}}}
{{1\over (8\psi)^{8n}}}\cr
&= ~{}_6 F_5 \left(
{\scr 1\over \scr 8},{\scr 2\over \scr 8},{\scr 3\over \scr 8},
{\scr 5\over \scr 8},{\scr 6\over \scr 8},{\scr 7\over \scr 8};\,
{\scr 1\over \scr 2},{\scr 1\over \scr 2},1,1,1;\, {1\over{2^6\psi^8}}
\right)~,}$$
which satisfies a sixth order hypergeometric differential equation.  One
may
check that precisely four of the indices of the equation corresponding to
$\psi\to\infty$ vanish. This means that four of the six solutions of the
equation behave like $1$, $\log\psi$, $\log^2\psi$ and $\log^3\psi$ as
$\psi\to\infty$. This also follows from the previously computed
Jordan form of the monodromy
about $D_{(-1,-1)}$, which consists of two blocks: there is a $4\times4$ block
with eigenvalue 1, and a $2\times2$ block with eigenvalue $-1$.
This fact will be of importance later when we come to
construct the mirror map.

For \Mtwo, the fundamental period turns out to be
$$
\vp_0(\ph,\ps) = \sum_{r,s = 0}^\infty
{(12r+6s)!\, (-2\ph)^s \over
(6r+s)! \left((2r+s)!\right)^2 (r!)^2\, s!\,
(12\ps)^{12r+6s}} \eqlabel{pi12}$$
When $\phi=0$, $\vp_0$ is given by
$$\varpi_0(\psi,0) =
{}~{}_6 F_5 \left(
{\scr 1\over \scr 12},{\scr 3\over \scr 12},{\scr 5\over \scr 12},
{\scr 7\over \scr 12},{\scr 9\over \scr 12},{\scr 11\over \scr 12};\,
{\scr 1\over \scr 2},{\scr 1\over \scr 2},1,1,1;\,
{1\over {2^{10} 3^6 \psi^{12}}} \right)~,$$
The six independent solutions of the associated hypergeometric
differential equation provide a basis of periods when $\phi=0$.
\subsection{The Picard--Fuchs equations}
The periods of the holomorphic three--form $\O$
can also be determined as solutions of the Picard--Fuchs
differential equations that are in fact equivalent to
the matrix equations discussed in Section~4.2.
As implied by eq. \eqref{mms}, each row of $R$ can be written
in terms of periods $\o$ (and their derivatives) that satisfy
a coupled system of equations.
For $\wP[8]$, eq. \eqref{mms} leads to the following system for
$\o\define\O/\ps$~:
$$\eqalign{
{\partial^3\o\over\partial\psi^3}
- 32\psi^3{\partial^3\o\over\partial\psi^2\partial\phi}
-96\psi^2{\partial^2\o\over\partial\psi\partial\phi}
-32\psi\pd{\o}{\phi} &= 0\cropen{15pt}
16(\phi^2-1){\partial^2\o\over\partial\phi^2}+
\psi^2{\partial^2\o\over\partial\psi^2}+
8\psi\phi{\partial^2\o\over\partial\psi\partial\phi}+
3\psi{\partial\o\over\partial\psi}+
24\phi{\partial\o\over\partial\phi} + \o &= 0\cropen{15pt}
(8\psi^4 + \phi + 1)(8\psi^4 + \phi - 1)
{\partial^4\o\over\partial\psi^3\partial\phi}
+128\psi^3(8\psi^4 + \phi)
{\partial^3\o\over\partial\psi^2\partial\phi}
+ 50\psi^3{\partial^2\o\over\partial\psi^2}
+\hskip5pt &\cr
+ 16\psi^2(148\psi^4 + 13\phi){\partial^2\o\over\partial\psi\partial\phi}
+ 30\psi^2\pd{\o}{\psi} +16\psi(44\psi^4 + 3\phi)\pd{\o}{\phi}
+2\psi\o &= 0~. \cr}\eqlabel{PFeqns}$$

This system is somewhat complicated and it is for this reason
that we base our development on the fundamental period $\vp_0$ that is
obtained by direct integration.
The equations \eqref{PFeqns} are however sometimes useful.
An example of this is provided by the fact that it is easy to show from
\eqref{PFeqns} that there are two periods which
in the limit that $\phi\to 1$
are given asymptotically by $(\phi - 1)^\half$ and
$(\phi - 1)^\half\log (\phi-1)$. A knowledge of these asymptotic forms is
useful
when discussing the degeneration of the manifold as $\phi\to 1$ and in
selecting an integer basis.
\subsection{Analytic properties of the fundamental period}
A useful alternative expression of the period \eqref{pi0double}
is obtained by setting $n=2r+s$ and summing over $n$ and $r$:
$$\eqalign{
\vp_0(\ph,\ps) &= \sum_{n=0}^\infty\sum_{r=0}^{\left[ {n\over2} \right]}
{(4n)!\,(-2\ph)^{n-2r}\over
(n!)^3\,(r!)^2\,(n-2r)!\,(8\ps)^{4n}}\cropen{5pt}
&=\sum_{n=0}^\infty{(4n)!(-1)^n \over (n!)^4 (8\psi)^{4n}}
\, u_n(\ph)~~,~~
\left|{\ph\pm 1\over 8\ps^4}\right|<1~,\cr}\eqlabel{pi0single}$$
where
$$
u_n(\ph) = (2\ph)^n\sum_{r=0}^{\left[ {n\over2}\right]}
{n!\over (r!)^2\, (n-2r)!\, (2\ph)^{2r}}\eqlabel{ydef}$$
and the convergence criterion follows from the
asymptotic form of $u_n$ for large $n$ which will be calculated presently.

The function $u_n(\ph)$ is a polynomial of degree
$n$. However it proves useful    to extend the definition of $u_n(\ph)$ to
complex values of $n$. To this end we note that we might as well take the
upper
limit of summation in \eqref{ydef} to be $\infty$ since the factor
$(n-2r)!=\G(n-2r+1)$ causes the summand to vanish for $r>\left[ {n\over2}
\right]$. By means of the duplication formula for the $\G$--function we
may
write
$$
{1\over \G(\n-2r+1)}= -2^{2r-\n-1}\p^{-{3\over 2}}\sin(\p \n)\,
\G\!\left( r-{\n\over 2}\right) \G\!\left( r-{\n\over 2}+{1\over 2}\right)
$$
In this manner we may relate $u_\n(\ph)$ to the hypergeometric function in
a
number of ways, the following relation being the most useful:
$$
u_\n(\ph) = (2\ph)^\n\,
\F{-{\n\over 2}}{-{\n\over 2}+{1\over2}}1{1\over\ph^2}~.\eqlabel{hgfs}$$
The erudite reader will recognise that these functions are
related to \hbox{Legendre} functions. However the
standard definitions take the branch cut to run from $\ph=-1$ to $\ph=+1$
which
would be awkward for our purposes since this would make $u_\n(\ph)$
discontinuous in a neighborhood of $\ph=0$. We choose instead to have the
cuts
run out along the real axis from the points $\ph=\pm 1$ to $\infty$. The
relation \eqref{hgfs} holds in the upper half--plane. Analytic
continuation of
$u_\n(\ph)$
to the rest of the plane is most easily accomplished by means of the
integral
representation
$$
u_\n(\ph)={2^\n \over \p}\int_{-1}^1{d\z\over\sqrt{1-\z^2}}\,(\ph-\z)^\n
\eqlabel{intrep}$$
which follows from Euler's integral for the hypergeometric function. It is
simplest to take \eqref{intrep} as the defining relation and we will use
this
as the basis of our study of the properties of $u_\n(\ph)$. Initially
$\ph$ is taken to lie in the upper half--plane. But we may analytically
continue to the lower half--plane (and even beyond, by crossing the
branch cut).  In so doing we must, of course, ensure that the contour of
integration in \eqref{intrep} is deformed ahead of $\ph$
as in figure 6.

{}From the integral representation a number of results follow easily. The
differential equation satisfied by $u_\n(\ph)$ is
$$
(\ph^2-1){d^2u_\n\over d\ph^2} - (2\n-1)\ph {du_\n\over d\ph} +
\n^2u_\n=0$$
whose second solution can be taken to be $u_\n(-\ph)$ though we shall see
that $\unu$ and $\unum$ are not linearly independent when $\n$ is an
integer
so we shall shortly introduce another solution.

An elementary but useful result is the value of $u_\n(\ph)$ at
$\ph=0$
$$u_\n(0)={2^\n \p^\half e^{i\p \n/ 2}
\over\G\left(1+{\n\over 2}\right)\G\left(\half-{\n\over 2}\right)}~.$$
In particular
$$
u_{2n+1}(0)=0~~~~~\hbox{and}~~~~~u_{-(2n+2)}(0)=0~~,~~~n=0,1,2,\ldots~.$$
More generally we have the relations
$$u_n(-\ph)=(-1)^n u_n(\ph) ~~~~~\hbox{and}~~~~~
u_{-n}(-\ph)=(-1)^{n+1}u_{-n}(\ph)~~,~~~n=0,1,2,\ldots~.
\eqlabel{integers}$$
These are elementary consequences of \eqref{intrep}. The second relation uses
also the well known fact that $P_\n (z)=P_{-\n-1}(z)$.

It is of interest to note with regard to the discussion of Section~3.2
concerning the degeneration of the manifold when $\ph=1$, that
$$
u_n(1)= {(2n)!\over (n!)^2}$$
as is easily seen from \eqref{intrep}. Thus we find that when $\ph=1$ the
fundamental period takes the form
$$
\vp_0(1,\ps) = \sum_{n=0}^\infty {(2n)!\,(4n)!\over (n!)^6}
{(-1)^n\over (8\ps)^{4n}}$$
which is the fundamental period for the mirror of the
manifold $\IP_5[2,4]$ (see for example \cite{\rperiods}).

Returning to the case of general $\ph$, it is also worth recording the
elementary fact that
$$
{du_\n(\ph)\over d\ph} = 2\n u_{\n-1}(\ph)~. \eqlabel{deriv}$$
{}From the integral representation we may derive the asymptotic forms,
$$\eqalign{
u_\n(\ph)&\asymp {2^{\n-\half}\over \sqrt{\p\n}}
\left\{ (\ph+1)^{\n+\half}\!\!
-\,\,i(\ph-1)^{\n+\half} \right\}~,\cr
u_\n(-\ph)&\asymp {2^{\n-\half}\over \sqrt{\p\n}}
\left\{ e^{i\p\n}(\ph+1)^{\n+\half}\!\!
-\,\,ie^{-i\p\n}(\ph-1)^{\n+\half} \right\}~,\cr}
{}~~~~~~\n\to\infty~,$$
we see that the the series \eqref{pi0double} and \eqref{pi0single}
converge
when
$$
\left|{\ph\pm 1\over 8\ps^4}\right|<1~.\eqlabel{bigpsi}$$

{}From the integral \eqref{intrep}, we find the
monodromy relations for the functions $u_\n(\ph)$ and $u_\n(-\ph)$. The
monodromy about $\ph=1$ is given by
$$\eqalign{
u_\n(\ph) &\to (1-e^{2i\p \n})u_\n(\ph) + e^{i\p \n}u_\n(-\ph)\cr
u_\n(-\ph)&\to e^{i\p \n}u_\n(\ph)~.\cr}
\eqlabel{mono}$$
While the monodromy about $\ph=-1$ is given by
$$\eqalign{
u_\n(\ph) &\to e^{i\p \n}u_\n(-\ph)\cr
u_\n(-\ph)&\to e^{i\p \n}u_\n(\ph) + (1-e^{2i\p \n})u_\n(-\ph)~.\cr}$$
We may also continue the functions around $\ph=\infty$. If we take
$|\ph|>1$,
replace $\ph$ by $\ph e^{i\th}$ and let $\th$ vary from 0 to $\p$ we find
$$\eqalign{
u_\n(\ph e^{i\p})  &=e^{i\n\p}\unu \cr
u_\n(-\ph e^{i\p}) &=(1-e^{2i\n\p})\unu + e^{i\n\p}\unum~.\cr}
$$

We have seen in \eqref{integers} that the functions $\unu$ and $\unum$
become
linearly dependent when $\n$ is an integer. In fact their Wronskian
\Footnote{Our convention is that $W[f,g]=fg'-f'g$.} has the form
$$
W[\unu,\unum] =
-{2^{2\n+1} \sin\n\p \over \p}(\ph^2 -1)^{\n-\half}~.$$
We therefore define
$$
\vnu = {\p\over\sin\n\p}\left(\unu\cos\n\p - \unum\right)~.$$
Since $W[u_\n,v_\n]=2^{2\n+1}(\ph^2 -1)^{\n-\half}$ the functions $u_\n$
and $v_\n$ are linearly independent for all $\n$. When $\n$ is an integer
we
have the relation
$$
\vn = \pd{\un}{\n} - (-1)^n\pd{\unm}{\n}~.\eqlabel{vn} $$
It turns out that in order to calculate the mirror map we shall also need
explicit expressions for the functions $\pd{\un}{\n}$. We therefore define
a
further set of functions for the case that $\n$ is an integer
$$
\wn = \pd{\un}{\n} + (-1)^n\pd{\unm}{\n}~.$$
We need now to find workable expressions for the $v_n$ and $w_n$. These
may be
found from the differential equation and the monodromy relations. The
$v_n$
transform under two operations. The first of these is $\ph\to -\ph$ (we
shall
call this operation $A$ since it is the restriction to $\ph$ of
$A:(\ps,\ph)\to(\a\ps,-\ph)$) and the second is $B$, transport around
$\ph=1$,
$$
\eqalign{
A:~\vn &\to (-1)^{n+1}\vn\cr
B:~\vn &\to -\vn - 2\p i\un~.\cr}$$
Consider first the case $n=0$. We have $u_0=1$ and it is easy in this case
to
to solve the differential equation directly. The solution with the
requisite
properties is
$$
v_0(\ph)= 2\log(\ph+\sqrt{\ph^2 - 1}) - i\p~.$$
For the case of general $n$ consider the function
$$
f_n(\ph)\define -{(\vn -\un v_0(\ph))\over\sqrt{\ph^2 - 1}}~.$$
The function $f_n$ is single valued under transport around $\ph=1$ hence
also
around $\ph=-1$ and around infinity. Thus $f_n$ contains no logarithms and
is
$O(\ph^{n-1})$ as $\ph\to\infty$. Since $f_n$ is, moreover, regular at
$\ph=\pm 1$ it is a polynomial. We conclude that
$$
\vn = \un v_0(\ph) - \sqrt{\ph^2 - 1}\fn \eqlabel{veq}$$
with $f_n$ a polynomial of order $(n-1)$.

Now by differentiating the differential equation with respect to $\n$ we
see
that
$$
\left\{ (\ph^2-1){d^2\over d\ph^2} - (2\n-1)\ph{d\over d\ph} +\n^2\right\}
\pd{u_\n}{\n} = 2\ph u^\prime_\n - 2\n u_\n ~.\eqlabel{diffnu}$$
{}From the integral representation \eqref{intrep} we have that
$$
\unu =(2\ph)^\n + O(\ph^{\n-2})~~\hbox{as}~~\ph\to\infty$$
The differential operator on the RHS of \eqref{diffnu} annihilates
$\ph^\n$ so
the RHS is in fact $O(\ph^{\n-2})$. Thus we may write
$$
\pd{\un}{\n} = A_n\un + B_n \vn + O(\ph^{n-2})$$
with the $O(\ph^{n-2})$ term a polynomial. By differentiating the integral
representation \eqref{intrep} and taking $\ph$ to be large we see that
$B_n=\half$ and there is a relation between $A_n$ and the leading term of
the
quantity $\ph\fn$. The conclusion is that
$$
\wn =i\p\un + \ph\fn + \gn \eqlabel{weq} $$
with $g_n$ a polynomial of order $(n-2)$.

The computation of the auxilliary polynomials $f_n$ and $g_n$ is
surprisingly
simple due to the relation \eqref{deriv} in virtue of which the
differential equation becomes a recurrence relation
$$
\n\unu=2(2\n-1)\ph u_{\n-1} - 4(\n-1)(\ph^2-1) u_{\n-2}\eqlabel{recur} $$
which, together with the initial values $u_0(\ph)=1~,~u_1(\ph)=2\ph$, is
an
efficient way to generate the $u_n$. By differentiating \eqref{recur} we
find
recurrence relations for the $v_n$ and the $w_n$ and these provide
recurrence
relations for the auxilliary polynomials. Thus we have
$$\eqalign{
n u_n&=2(2n-1)\ph u_{n-1} -4(n-1)(\ph^2-1)u_{n-2}~,\cropen{5pt}
n f_n&=2(2n-1)\ph f_{n-1} -4(n-1)(\ph^2-1)f_{n-2}~,\cropen{5pt}
n g_n&=2(2n-1)\ph g_{n-1} -4(n-1)(\ph^2-1)g_{n-2}\cr
&\kern50pt - 2 u_n + 8\ph u_{n-1} -
8 (\ph^2-1)u_{n-2}~,\cr}
\kern65pt
\eqalign{
u_0&=1~,~u_1=2\ph~,\cropen{5pt}
f_0&=0~,~f_1=4~,\cropen{5pt}
   &\cr
g_0&=0~,~g_1=0~.\cr}
$$
The first few polynomials are presented in Table~\tabref{polys}.
\vfill
\def\shortrule{\vrule height 12pt depth 8pt}
\def\endrule{\vrule height5pt}
$$\vbox{\offinterlineskip\halign{
#& \hskip30pt$#$\hskip30pt\hfil\vrule&
\hskip25pt$#$\hskip50pt\hfil\vrule\cr
\noalign{\hrule}
\endrule &$$ $$ &$$ $$\cr
\shortrule &$$ u_0 = 1$$ &$$ $$\cr
\shortrule &$$ u_1 = 2 \phi $$ &$$ $$\cr
\shortrule &$$ u_2 = 2 + 4 {{\phi }^2}$$ &$$ $$\cr
\shortrule &$$ u_3 = 12 \phi  + 8 {{\phi }^3}$$ &$$ $$\cr
\shortrule &$$ u_4 = 6 + 48 {{\phi }^2} + 16 {{\phi }^4}$$ &$$ $$\cr
\shortrule &$$ u_5 = 60 \phi  + 160 {{\phi }^3} + 32 {{\phi }^5}$$ &$$
$$\cr
\shortrule &$$ u_6 = 20 + 360 {{\phi }^2} + 480 {{\phi }^4} +
64 {{\phi }^6}$$ &$$ $$\cr
\endrule &$$ $$ &$$ $$ \cr
\noalign{\hrule}
\endrule &$$ $$  & $$ $$\cr
\shortrule &$$ f_0 = 0$$
&$$ g_0 = 0$$  \cr
\shortrule &$$ f_1 = 4$$
&$$ g_1 = 0$$ \cr
\shortrule &$$ f_2 = 12 \phi $$
&$$ g_2 = 2$$ \cr
\shortrule &$$ f_3 = {{32}\over 3} + {{88 {{\phi }^2}}\over 3} $$
&$$ g_3 = {{28 \phi }\over 3}$$ \cr
\shortrule &$$ f_4 = {{220 \phi }\over 3} + {{200 {{\phi }^3}}\over 3}$$
&$$ g_4 = 7 + {{92 {{\phi }^2}}\over 3}$$ \cr
\shortrule &$$ f_5 = {{512}\over {15}} + {{4856 {{\phi }^2}}\over {15}} +
{{2192 {{\phi }^4}}\over {15}}$$
&$$ g_5 = {{898 \phi }\over {15}} +
{{1304 {{\phi }^3}}\over {15}}$$ \cr
\shortrule &$$ f_6 = {{1848 \phi }\over 5} +
{{5824 {{\phi }^3}}\over 5} + {{1568 {{\phi }^5}}\over 5}$$
&$$ g_6 = {{74}\over 3} +
{{1572 {{\phi }^2}}\over 5} + {{1136 {{\phi }^4}}\over 5} $$ \cr
\endrule &$$ $$ &$$ $$\cr
\noalign{\hrule}
}}
$$
\nobreak\tablecaption{polys}{The first few polynomials $u_n$ together with
the auxilliary polynomials $f_n$ and $g_n$.}
%
\newpage
For future purposes we need to introduce the combination
$h_n = \ph f_n + g_n$. It can be shown that
$$ h_n(\ph) = 2(2\phi)^n  \sum_{k=0}^{\left[ {n\over 2}\right]}
{ n!\over {(k!)^2 (n-2k)! (2\phi) ^k} }
[ \Ps(1+n) - \Ps(1+k) ]~, $$
where $\Psi(z)$ is the Digamma function.

The results for the function $u_n(\ph)$ are also applicable to the
\Mtwo \ model. Indeed, in this case
the fundamental period can be cast in the form
$$
\vp_0(\ph,\ps) =
\sum_{n=0}^\infty{(6n)! (-1)^n \over (n!)^3 (3n)! (12\psi)^{6n}}
\, u_n(\ph)~~,~~
\left|{\ph\pm 1\over 864\,\ps^6}\right|<1 ~, \eqlabel{pu12}$$
where $u_n(\ph)$ is precisely the same function as that given in eq.
\eqref{ydef}.
When $\ph=1$, $\vp_0$ becomes the fundamental period of the
mirror of $\IP_5^{(1,1,1,1,1,3)}[2,6]$ (as calculated in \cite{\rperiods}).

The above considerations will shortly enable us to compute the mirror
map. First however we need to find explicit expressions for the
periods. It is to this that we now turn.
%
%
\subsection{Analytic continuation of the periods}
In order to obtain an expression for $\vp_0$ valid for small $\ps$ we
proceed as in \cite{\rCdGP}. We will only show detailed results for \Mone.
The first step is to write the period as an integral of Barne's type
$$
\vp_0(\ps,\ph) = {1\over 2\p i}\int_C d\n
{\G(-\n)\G(1+4\n)\over\G^3(1+\n)}
(2^{12}\ps^4)^{-\n}u_\n(\ph)~,~~~~
-\p < \arg\left({8\ps^4\over\ph\pm 1}\right) < \p~.$$
If the inequality \eqref{bigpsi} is satisfied then the contour can be
closed to
the right encompassing the poles of $\G(-\n)$ and \eqref{pi0single} is
recovered as a sum of residues. If, on the other hand $|8\ps^4|<|\ph\pm
1|$
then the contour may be closed to the left encompassing the poles of
$\G(1+4\n)$. In  this
way we obtain an expression for $\vp_0$ and by acting on this with
$\ca{A}^j$
we find
$$
\vp_j(\ps,\ph)= - {1\over 4}\sum_{m=1}^\infty{(-1)^m\a^{mj}\G({m\over
4})\over
\G(m)\G^3(1-{m\over 4})}(2^{12}\ps^4)^{m\over 4}
u_{-{m\over 4}}((-1)^j\ph)~~~,~~~\left|{8\ps^4\over \ph\pm 1}\right|<1~.
\eqlabel{pismall}$$
We now have explicit expressions for the $\vp_j$. Note that the terms
containing $\ps^{4n}$ are in fact absent owing to the factors of
$\G(1-{m\over 4})$. Thus we have
$$\eqalign{\vp_0+\vp_2+\vp_4+\vp_6&=0\cr
           \vp_1+\vp_3+\vp_5+\vp_7&=0\cr}$$
expressing the fact that only six of the $\vp_j$ are linearly independent.

We wish to compute the monodromy of these periods around the singularity
at $\ph=1$ using the relations \eqref{mono}. We cannot however do this
directly since the process of analytic continuation about $\ph=1$ takes us
out
of the region of convergence of \eqref{pismall}. We must therefore first
continue the $\vp_j$ to large values of $\ps$ and then use \eqref{mono}.
To
this end we write the variable of summation in \eqref{pismall} as $4n+r$.
We
have
$$\eqalign{
\vp_{2j} &= -{1\over 4\p^3}\sum_{r=1}^3 (-1)^r \sin^3\left({\p r\over
4}\right)
\a^{2jr}\x_r\cr
\vp_{2j+1} &= -{1\over 4\p^3}\sum_{r=1}^3(-1)^r\sin^3\left({\p
r\over4}\right)
\a^{(2j+1)r}\eta_r~,\cr}\eqlabel{mnint}$$
with
$$\eqalign{
\x_r(\ps,\ph)&=\sum_{n=0}^\infty
{\G^4(n+{r\over 4})\over\G(4n+r)}(2^{12}\ps^4)^{n+{r\over 4}}
(-1)^n u_{-(n+{r\over 4})}(\ph)~,\cr
\eta_r(\ps,\ph)&=\sum_{n=0}^\infty
{\G^4(n+{r\over 4})\over\G(4n+r)}(2^{12}\ps^4)^{n+{r\over 4}}
u_{-(n+{r\over 4})}(-\ph)~.\cr}\eqlabel{munu}$$
The only difference between the expressions for $\x_r$ and $\eta_r$ is
the
replacement of\break $(-1)^n u_{-(n+{r\over 4})}(\ph)$ by
$u_{-(n+{r\over 4})}(-\ph)$. The point of this exercise is that it is
easier
to continue the $\x_r$ and the $\eta_r$ than the $\vp_j$. The continuation
of
these new functions is again accomplished by introducing integral
representations:
$$\eqalign{
\x_r(\ps,\ph) &=
\- \int_C {d\n\over 2i \sin\p(\n+{r\over 4})}{\G^4(-\n)\over\G(-4\n)}
(2^{12}\ps^4)^{-\n}u_\n(\ph)\cropen{10pt}
\eta_r(\ps,\ph) &
= - \int_C {d\n\over 2i \sin\p(\n+{r\over 4})}{\G^4(-\n)\over\G(-4\n)}
(2^{12}\ps^4)^{-\n}{\left[ u_\n(\ph)
\sin\p(\n+{r\over 4}) + u_\n(-\ph)\sin({\p r\over 4})\right]\over
\sin\p\n}~.\cr}\eqlabel{xieta}$$
Again these representations are justified by closing the
contours to the left when $\ps$ is small and checking that the previous
expressions\eqref{munu} are recovered as sums over the residues associated
with
the zeros of $\sin\p(\n+{r\over 4})$. Note that, despite appearances, the
factor
of $\sin\p \n$ in the $\eta_r$ integral does not introduce additional
poles on
the negative real axis since the quantity in the numerator vanishes for
$\n = -1,-2,\ldots$ in virtue of \eqref{integers}. The $\eta_r$ integrand
is
more
complicated than the $\x_r$ integrand but has to be taken this way in
order to
secure convergence of the integral. When $\ps$ is large the contours can
be
closed to the right so as to encompass the poles where $\n$ is an integer.
These
are of third order for $\x_r$ and of fourth order for $\eta_r$.

Finally we are able to compute the monodromy of the periods about $\ph=1$.
For
this it suffices to consider just the contributions of the poles at $\n=0$
and
$\n=1$ and to set $\ph=0$. Setting $\ph=0$ changes the order of the poles
in
virtue of the fact that $u_\n(0)$ has zeros at the positive integers. The
assiduous reader will check that $\x_r$ has a triple pole at $\n=0$ and a
double
pole at $\n=1$ while $\eta_r$ has a fourth order pole at $\n=0$ and a
double
pole at $\n=1$. The residues at these poles are linear combinations of the
six
quantities
$$1~,~\log\ps~,~\log^2\ps~,~\log^3\ps~,~\ps^{-4}~,~\ps^{-4}\log\ps~,$$
as was anticipated on the basis of our analysis of the periods for
$\ph=0$, as well as by the calculation of the Jordan form of the monodromy
about $D_{(-1,-1)}$.
By employing the rules \eqref{mono}, again setting $\ph=0$ and calculating
the new residues we find
$$
\vp\to \sB\vp$$
where
$$
\vp = \pmatrix{\vp_0\cr \vp_1\cr \vp_2\cr \vp_3\cr \vp_4\cr
\vp_5\cr}~~~~~~
\hbox{and}~~~~~~
\sB = \matrixb~.$$

The computation of the monodromy about the conifold is straightforward.
The
periods have the structure
$$
\vp_j(\ps,\ph) = {c_j\over 2\p i}\ca{G}_0\,\log(8\ps^4 - \ph - 1) +
f_j(\ps,\ph)$$
where $\ca{G}_0$, the period corresponding to the 3--sphere that vanishes at
the conifold, and the $f_j$ are functions that
are analytic in a neighborhood of the conifold.
For the purpose of computing the coefficients $c_j$ it suffices
to again set $\ph=0$ and to observe that the coefficient
of the logarithm can be found by studying the convergence of
the series \eqref{pismall}. If $\ca{G}_0$ has the asymptotic form
$$\ca{G}_0(\ps,0)\asymp \hbox{const.} (8\ps^4 - 1)$$
then
$$
{d^2 \vp_j\over d\ps^2}\asymp \hbox{const.} {c_j\over (8\ps^4 - 1)}~.$$
On the other hand the leading behaviour of
the series \eqref{munu} can be found by use of Stirling's formula
$$
\G(z)\asymp \sqrt{2\p}z^{z-\half}e^{-z}~.$$
There is an overall multiplicative constant that needs to be fixed.
However we
know also, by means of an argument precisely parallel to that made in
\cite{\rCdGP} that $\vp_1\to \vp_0$ and
$$
\ca{G}_0 = \vp_1 -\vp_0~.$$
Thus we find
$$
\T :~\vp\to \sT\vp~~~~\hbox{with}~~~~
\sT = \matrixt~.$$
As expected, the Jordan form of $\sT$ is indeed the exponential of
\eqref{jordan}.

For completeness we record the matrix corresponding to monodromy about
$C_0$ which is the operation \hbox{$\ca{A}:~(\ps,\ph)\to (\a\ps,-\ph)$}
$$
\ca{A}:~\vp\to \sA\vp~~~~\hbox{with}~~~~
\sA = \matrixa~.$$
We also calculate a monodromy transformation
$\ca{T}_\infty$, the composite of $\ca{A}^{-1}$ and $\T^{-1}$,
which will be useful later;
its matrix is
$$
\sT_\infty=(\sA\sT)^{-1}=\matrixtinf~.$$
\newpage
\section{flatcoords}{The Mirror Map and Large Complex Structure Limit}
\vskip-20pt
\subsection{Generalities}
We wish to find the explicit map between the (extended) \K-cone of \M\ and
the
space of complex structures of \W. Both of these spaces enjoy special
geometry.
The mirror map follows by relating the period vectors for these two
spaces.

Now for the complex structures we have a period vector
$$
\P=\pmatrix{\ca{G}_0\cr \ca{G}_1\cr \ca{G}_2\cropen{3pt}
z^0\cr z^1\cr z^2\cr}~,\qquad \ca{G}_a = \pd{\ca{G}}{z^a}\eqlabel{Pidef}$$
such that the new periods correspond to a basis that is integral and
symplectic. In other words we need to find a homology basis
$(A^a,B_b),~a,b=0,1,2$ with
$$
A^a\cap A^b=0~~,~~B_a\cap B_b=0~~,~~A^a\cap B_b=\d^a_b~.$$
The components of $\P$ are then given by
$$
z^a=\int_{A^a}\O~~~~,~~~~\ca{G}_a=\int_{B_a}\O~.$$
We may choose $A^0$ to be the torus corresponding to our
fundamental period $\vp_0$ and $B_0$ to be the three--sphere that shrinks
to
zero at the conifold. Thus
$$
z^0=\vp_0~~\hbox{and}~~\ca{G}_0=\vp_1 - \vp_0~.$$
The argument that $A^0$ and $B_0$ meet in a single point precisely
parallels an
argument given in \cite{\rCdGP}.
For a given choice of symplectic basis $(A^a,B_b)$ there will be a
constant
real matrix $m$ such that
$$
\P=m\vp~.$$
On the \K\ side the analogue of the decomposition into $A$ and $B$ cycles
is
the decomposition of a period vector
$$
\ip = \pmatrix{\ca{F}_0\cr \ca{F}_1\cr \ca{F}_2\cropen{3pt}
w^0\cr w^1\cr w^2\cr}~,\qquad  \ca{F}_a =
\pd{\ca{F}}{w^a}\eqlabel{ipdef}$$
with respect to $H^0{\oplus} H^2{\oplus} H^4{\oplus} H^6$. The generators
of
$H^0$ and $H^6$ are special and we identify them with $A^0$ and $B_0$.
The flat structure here is identified with the natural flat structure on
the
\K-cone
$$
B+iJ~=~t^j e_j\eqlabel{flatcds}$$
where the $e_j$ are a basis for $H^2(\M,\IZ)$ and $t^j=w^j/w^0$. For the
case
in hand we may take $e_j=(H,L)$.
\subsection{The large complex structure limit for \Mone}
We have earlier identified the location of the large complex structure
limit, both by using monodromy properties and by using the monomial divisor
mirror map.  Now we wish to study the monodromy properties more carefully,
in order to find the correct coordinates on the moduli space.

Recall the form of the monodromy properties \eqref{lcsl}:
if the $S_i$ represent
monodromy transformations about boundary divisors, and we set
$R_i=S_i-{\bf 1}$, then we have
$$
\llap{\hbox{$\eqalign{i.&\cr ii.&\cr iii.&\cr}$\hskip50pt}}
\eqalign{\left[ R_i,R_j \right] &= 0\cr
R_iR_jR_k &= \yzero_{ijk}Y\cr
R_iR_jR_kR_l &= 0\cr}\eqlabel{lcslbis}$$
where $Y$ is a matrix independent of $i$.

Let us now consider our \Mone \ model in detail.
We examine the intersection of $D_{(0,-1)}$ and $C_\infty=D_{(1,0)}$.
The monodromies of
$\vp$ about these curves are

$$
\sC_\infty = (\sA\sT\sB)^{-1} = \sB^{-1}\sT_\infty~~~~,~~~~
\sD_{(0,-1)} = \sT_\infty^2~.$$
We set
$$
\sR_1 = \sT_\infty^2 - {\bf 1}~~~~,~~~~\sR_2 = \sB^{-1}\sT_\infty  - {\bf
1}
$$
and we immediately check that
$$
\left[ \sR_1 , \sR_2 \right] = 0~~,~~ \sR_1^3 =8\sY~~,~~
\sR_1^2\sR_2 = 4\sY~~,~~\sR_2^2=0~$$
with $\sY$ a certain matrix. Moreover
$$
\sR_1 \sY = \sR_2 \sY = 0$$
so that we see that $\sR_1$ and $\sR_2$ have the same algebra as $H$
and $L$ and we conclude once again
that $D_{(0,-1)}\cap C_\infty$ corresponds to the \lcsl.

We now wish to relate explicitly the $w$ basis to the $\vp$ basis.
We may regard the cycles as corresponding to row vectors such that their
inner
product with
the period vector $\vp$ gives the corresponding period. Thus, for example,
$$
A^0=(1, 0, 0, 0, 0, 0)~~~\hbox{and}~~~
B_0=( -1, 1, 0, 0, 0, 0)$$
since
$$
w^0=A^0\vp=\vp_0~~~\hbox{and}~~~\ca{F}_0=B_0\vp=\vp_1 - \vp_0~.$$

The identification of the $t^i$ follows by elementary considerations of
linear
algebra. We will have
$$
t^i = {A^i \vp \over \vp_0}~~~~,~~~~i=1,2$$
with the $A^i$ two row vectors. The effect of monodromy $\cS_j$ on $\vp$
is
$\vp\to \sS_j\vp$ or
equivalently $A^i\to A^i \sS_j$. in terms of the \sR\ matrices this
amounts to
$$
A^i\sR_j~=~A^0\,\d^i{}_j~.$$
This determines the $A^i$ up to the addition of a multiple of $A^0$ \ie\ $t^i$
is determined up to the addition of a constant.
Applying these general considerations to the model at hand, and with a certain
choice of constants that simplify later expressions, we find
$$
A^1 = {1\over 4}(-1,0,2,0,1,0)~~~,~~~A^2= {1\over 4}(1,3,-2,2,-1,1)$$
and hence $t^1$ and $t^2$ are given by
$$\eqalign{
t^1&=  -{1\over 4} + {(2\vp_2 + \vp_4)\over 4\vp_0}\cr
t^2&=\- {1\over 4} - {(2\vp_2 + \vp_4)\over 4\vp_0}
+ {(3\vp_1 + 2\vp_3 + \vp_5)\over 4\vp_0}~.\cr}\eqlabel{mmap}$$
These relations, to which we shall shortly return, express the mirror map.
First however we shall complete the process of finding the symplectic
basis.

We now know the four quantities $\ca{F}_0$, $w^0$, $w^1$ and $w^2$
in terms of the $\vp_j$ and we need to find similar expressions for the
two remaining periods $\ca{F}_1$, $\ca{F}_2$. To this end we set
$$\eqalign{
\ca{F}&= -{1\over 6}\bigg( 8\, (t^1)^3 + 3\cdot 4\, (t^1)^2 t^2\bigg)
+\half\bigg(\a\, (t^1)^2 + 2\b\, t^1 t^2 + \g\, (t^2)^2\bigg)\cr
&\hskip50pt +\bigg((\d-{\txt{2\over 3}})\, t^1 + \e\, t^2\bigg) +
\x + \cdots~~.\cr}\eqlabel{prepotF}$$
The constants $(\a,\b,\ldots,\e)$ can be adjusted by symplectic
transformations. The matrix
$$
  \pmatrix{{\bf 1}& Q\cr
           {\bf 0}& {\bf 1}\cr}~~~~,~~~~
Q=\pmatrix{       0&      \d'&     \e'\cr
                \d'&      \a'&     \b'\cr
                \e'&      \b'&     \g'\cr}\eqlabel{Sp}$$
is symplectic and shifts the constants $\a \to \a + \a'$ etc. From the
prepotential \eqref{prepotF} we find matrices $S_1$, $S_2$ corresponding
to
the monodromy $\cS_i$. We identify $\ip$ with $\P$ so
$$
\ip~=~\P~=~ m\vp\eqlabel{ident}$$
the matrix $m$ has the vectors $(A^a,B_b)$ as its rows and hence is of the
form
$$
m=\pmatrix{
            -1&     1&     0&     0&     0&     0\cropen{3pt}
        \- y_0&\- y_1&\- y_2&\- y_3&\- y_4&\- y_5\cropen{3pt}
        \- z_0&\- z_1&\- z_2&\- z_3&\- z_4&\- z_5\cropen{3pt}
        \-   1&\-   0&\-   0&\-   0&\-   0&\-   0\cropen{3pt}
   -{1\over 4}&\-   0&\-  {1\over 2}&\- 0&\- {1\over 4}&\- 0\cropen{3pt}
\-{1\over 4}&\-{3\over 4}&-{1\over 2}&\-{1\over 2}&-{1\over 4}&\-{1\over
4}\cr}
{}~.\eqlabel{mdef}$$
The point is that the prepotential \eqref{prepotF}\ and equations
\eqref{ident}
and \eqref{mdef} are consistent only for a choice of the undetermined
parameters that is unique up to to the action of ${\rm Sp}(6,\IZ)$.
To show this explicitly we start with the prepotential \eqref{prepotF} and
consider the effect of making, in turn, the replacements $t^1\to t^1+1$
and
$t^2\to t^2+1$. In this way we find matrices
$$
\def\hide#1{\hidewidth{#1}\kern-5pt}
S_1=\pmatrix{
1 &  -1 &\- 0 & 2\,\delta    &\hide{\alpha{+}4}  &\hide{\beta{+}2}\-  \cr
0 &\- 1 &\- 0 &\hide{\alpha{-}4} & -8            &  -4\- \cr
0 &\- 0 &\- 1 &\hide{\beta {-}2}   & -4          &\- 0\- \cr
0 &\- 0 &\- 0 &\- 1            &\- 0             &\- 0\- \cr
0 &\- 0 &\- 0 &\- 1            &\- 1             &\- 0\- \cr
0 &\- 0 &\- 0 &\- 0            &\- 0             &\- 1\- \cr  }~~,~~
S_2=\pmatrix{
1 &\- 0 &  -1 &\- 2\,\e  &\- \beta  &\- \gamma  \cr
0 &\- 1 &\- 0 &\- \beta  & -4       &\- 0 \cr
0 &\- 0 &\- 1 &\- \gamma &\- 0      &\- 0 \cr
0 &\- 0 &\- 0 &\- 1      &\- 0      &\- 0 \cr
0 &\- 0 &\- 0 &\- 0      &\- 1      &\- 0 \cr
0 &\- 0 &\- 0 &\- 1      &\- 0      &\- 1 \cr }~.
$$
The matrices $S_i$ and $\sS_i$ must be related by $S_i = m\sS_i m^{-1}$ or
equivalently
$$
S_i m - m \sS_i = 0~~,~~i=1,2~.\eqlabel{mconds}$$
These conditions lead to a number of linear equations for the unknowns
$(y_j,z_k)$ in $m$. (It is a useful fact that the implementation of the
conditions \eqref{mconds} does not require the inversion of $m$.) Solving
these equations determines all of the $(y_j,z_k)$ as linear combinations
of the
parameters $(\a,\b,\ldots,\e)$. The next step is to impose the condition
that
the monodromy matrix $T=m\sT m^{-1}$ be symplectic. This determines $\d$
and
$\e$
$$
\d=-3~~~~,~~~~\e=-1~.$$
In fact we find
$$
m\sT m^{-1}=\matrixTF$$
as was to be expected. It remains to find the parameters $\a,\b,\g$. The
matrices $A=m\sA m^{-1}$ and $B=m\sB m^{-1}$ must also be integral and
symplectic. They are symplectic for all values of the parameters but they
are
both integral only if $\a,\b,\g$ are all integral. Being integral these
three
parameters can be set to any value by means of an ${\rm Sp}(6,\IZ)$
transformation of the form \eqref{Sp}. Perhaps the simplest choice is to
set
$$
\a=0~,~~~\b=-2~,~~~\g=0~.$$
This yields the following value for the matrix $m$
$$
m=\pmatrix{
 -1           &\-  1 &\-  0          &\-  0 &\-  0          &\-  0
\cropen{3pt}
\- 1          &\-  0 &\-  1          &   -1 &\-  0          &   -1
\cropen{3pt}
\- {3\over 2} &\-  0 &\-  0          &\-  0 & -{1\over 2}   &\-  0
\cropen{3pt}
\- 1          &\-  0 &\-  0          &\-  0 &\-  0          &\-  0
\cropen{3pt}
-{1\over 4}   &\-  0 &\-  {1\over 2} &\-  0 &\-  {1\over 4} &\-  0
\cropen{3pt}
\- {1\over 4}&\- {3\over 4}&-{1\over 2}&\- {1\over 2}&-{1\over 4}
&\-{1\over4}\cr  }~.$$
\subsection{Inversion of the mirror map}
We return now to the mirror map \eqref{mmap}. Our task is to invert this
relation to obtain $\ps(t^1,t^2)$ and $\ph(t^1,t^2)$.
It is convenient, in this case, to make the change of variable
$$\eqalign{
t=2\,t^1 + t^2 &=-\half + {1\over 4\vp_0}\left[(\vp_0 + 2\vp_2 +\vp_4) +
(3\vp_1 + 2\vp_3 + \vp_5)\right]\cropen{7pt}
s=\hphantom{2\,t^1} - t^2 &=-\half + {1\over 4\vp_0}\left[(\vp_0 + 2\vp_2
+\vp_4) -
(3\vp_1 + 2\vp_3 + \vp_5)\right]~.\cr}\eqlabel{ts}$$

We have integral representations \eqref{mnint} and \eqref{munu}
for the $\vp_j$ and these quantities simplify considerably when we form
the
combinations that we need. We find
$$
\left. \matrix{t\cr s\cr}\right\} =
-\half + {1\over 4\vp_0}\int_C {d\n \over\sin^2 \n\p}
{\G(1+4\n)\over \G^4(1+\n)}(2^{12}\ps^4)^{-\n}
(\unu\cos\n\p \pm \unum)~.$$
The higher order poles have cancelled and we have now second order poles
for
$t$ and first order poles for $s$. Note that the residues for $t$ involve
the
functions $w_n$ while those for $s$ involve the $v_n$.
In this way we find
$$
\eqalign{
i\p(t-1) &= \log\left({1\over (8\ps)^4}\right) + {1\over 2\vp_0}
\sum_{n=1}^\infty {(4n)!\over (n!)^4} {(-1)^n\over (8\ps)^{4n}}
\Bigl[ 2 A_n u_n + h_n\Bigr] \cropen{8pt}
i\p s &= \cosh^{-1}\ph - {\sqrt{\ph^2-1}\over 2\vp_0}
\sum_{n=1}^\infty {(4n)!\over (n!)^4} {(-1)^n\over (8\ps)^{4n}} f_n\cr}
\eqlabel{tseqs}$$
where the relations \eqref{veq} and \eqref{weq} have been
used. We have also defined the coefficient
\hbox{$A_n=4[\Psi(4n+1) - \Psi(n+1)]$} and have recognised that
$$
\log(\ph + \sqrt{\ph^2-1}) = \cosh^{-1}\ph~.$$
We exponentiate the first of \eqref{tseqs} and take the $\cosh$ of the
second in order to obtain equations suitable for iterative solution:
$$\eqalign{
{1\over (8\ps)^4} &= - q^\half \exp\left\{-{1\over 2\vp_0}
\sum_{n=1}^\infty {(4n)!\over (n!)^4} {(-1)^n\over (8\ps)^{4n}}
\Bigl[ 2 A_n u_n + h_n \Bigr]\right\}
\cropen{12pt}
\ph &= \cos\left\{\p s + {\sqrt{1-\ph^2}\over 2\vp_0}
\sum_{n=1}^\infty {(4n)!\over (n!)^4} {(-1)^n\over (8\ps)^{4n}}f_n\right\}
\cr} \eqlabel{iter}$$
where we have introduced $q\define e^{2\p i t}$.
The zero'th order solutions valid in the limit of large radius,
\ie\ large $\imag\, t$, are
$$
{1\over (8\ps)^4} \asymp - q^\half~~~~\hbox{and}~~~~\ph \asymp \cos \p
s~.$$
Inserting these expressions into the RHS of \eqref{iter} we obtain first
order
expressions and so on. In this way we find
$$\eqalign{
{1\over (8\ps)^4} &=
-q^\half + 256 \cos (\pi  s) {q} -
\Bigl( 2256 + 17040 \cos (2 \pi  s) \Bigr)  {q^{3\over 2}} + \cr
&\kern40pt\Bigl( 1963520 \cos (\pi  s) + 872960 \cos (3 \pi  s)
\Bigr){q^2} +
\cdots\cropen{10pt}
\ph &=
\cos (\pi  s) - \Bigl( 24 - 24 \cos (2 \pi  s) \Bigr)  q^\half -
\Bigl( 996 \cos (\pi  s) - 996 \cos (3 \pi  s) \Bigr) {q} -\cr
&\kern40pt \Bigl( 213120 - 165376 \cos (2 \pi  s) +
     47744 \cos (4 \pi  s) \Bigr)  {q^{3\over 2}} + \cdots\cr}$$

In this example it has proved possible to expand in series for $q$
small and fixed $s$. As a matter of practical computation this has
some advantages. However, this procedure may not generalize to other
cases and we are, in general, obliged to
resort to a double series in the variables $q_i = e^{2\pi i t_i}$.
It is interesting then to recast the mirror map in terms of the
variables $(t^1,t^2)$. To this end we introduce `large
complex structure coordinates'
$$Z_1 = { {(8\ps)^4}\over {2\ph} }   \quad ; \quad  Z_2 = (2\ph)^2 ~.$$
Note that these are, up to multiplication by constants, the inverses of the
coordinates given in
Table~\tabref{fourregs1}.  Inverses are used to ensure that the coordinates
become large in the \lcsl.
{}From previous results we obtain
$$\eqalign{
2i\pi t^1 & = i\pi  -\log Z_1 - \sum_{n=1}^{\infty} \,
{{(2n-1)!}\over {(n!)^2}} \, Z_2^{-n} \cropen{10pt}
&+ {1\over {2\vp_0}} \, \sum_{n=1}^{\infty} \,
{{(4n)! (-1)^n}\over {(n!)^4}} \, Z_1^{-n} [2A_n\hat u_n(Z_2)
+ 2\hat h_n(Z_2) + \hat f_n(Z_2) ]~, \cropen{18pt}
2i\pi t^2 & = -\log Z_2 + 2 \sum_{n=1}^{\infty} \,
{{(2n-1)!}\over {(n!)^2}} \, Z_2^{-n} \, -
{1\over {\vp_0}} \, \sum_{n=1}^{\infty} \,
{{(4n)! (-1)^n}\over {(n!)^4}} \, Z_1^{-n} \hat f_n (Z_2)~,  \cr}$$
where we have defined
$$\eqalign{
u_n &= (2\ph)^n \hat u_n \quad ; \quad
h_n = 2 (2\ph)^n \hat h_n~,  \cropen{10pt}
f_n &= - {(2\ph)^n\over {\sqrt{\ph^2 -1}} } \hat f_n~. \cr}
\eqlabel{uhfhat}$$
We have also used the result
$$i \cos^{-1} \ph = \log (2\ph) - \sum_{n=1}^{\infty} \,
{{(2n-1)!}\over {(n!)^2}} (2\ph)^{-2n}~. $$
Notice that the large radius limit $\imag\, \ t^i \to \infty$
manifestly corresponds to the large complex structure limit
$Z_i \to \infty$.

The mirror map for \Mtwo \ follows \hbox{\it mutatis mutandis}.
In this model we choose to expand in a double series in terms of the large
complex structure coordinates which, for this case, are given by
$$Y_1 = { {(12\ps)^6}\over {2\ph} }   \quad ; \quad  Y_2 = (2\ph)^2 ~.$$
We then have
$$\eqalign{
2i\pi t^1 & = i\pi  -\log Y_1 - \sum_{n=1}^{\infty} \,
{{(2n-1)!}\over {(n!)^2}} \, Y_2^{-n} \cropen{10pt}
&+ {1\over {2\vp_0}} \, \sum_{n=1}^{\infty} \,
{{(6n)! (-1)^n}\over {(n!)^3 (3n)!}} \, Y_1^{-n} [2B_n\hat u_n(Y_2)
+ 2\hat h_n(Y_2) + \hat f_n(Y_2) ]~, \cropen{18pt}
2i\pi t^2 & = -\log Y_2 + 2 \sum_{n=1}^{\infty} \,
{{(2n-1)!}\over {(n!)^2}} \, Y_2^{-n} \, -
{1\over {\vp_0}} \, \sum_{n=1}^{\infty} \,
{{(6n)! (-1)^n}\over {(n!)^3 (3n)!}}\, Y_1^{-n} \hat f_n (Y_2)~.  \cr}$$
Here, $B_n=6\Psi(6n+1) - 3\Psi(3n+1) - 3\Psi(n+1)$, where
$\hat u_n, \hat h_n$ and $\hat f_n$ are the same functions as defined
in \eqref{uhfhat}. The fundamental period is given in \eqref{pu12}.

Inverting the mirror map yields expansions for the automorphic
functions $Y_1$ and $Y_2$ in terms of the variables
$q_i$. For instance, to third order
$$\eqalign{
Y_1 &=-{1\over q_1}(1 + 744q_1 - q_2
+ 196884q_1^2 + 480q_1q_2 + q_2^2\cropen{10pt}
&\hskip30pt+ 21493760q_1^3
+ 1403748q_1^2q_2 - 960q_1q_2^2 - q_2^3 + \cdots)~, \cropen{15pt}
Y_2 &= {1\over q_2}(1 + 240q_1 + 2q_2 + 70920q_1^2 - 240q_1q_2 +
q_2^2 \cropen{10pt}
&\hskip30pt+ 22696640q_1^3 - 57600 q_1^2q_2 - 240 q_1q_2^2 + \cdots)~.
  \cr}$$

As a curiosity, note that $Y_1$ restricts to the negative of the $j$~invariant
on the locus $q_2=0$.
\newpage
\section{yukawas}{The Yukawa Couplings and the Instanton Expansion}
\vskip-20pt
\subsection{The couplings}
The four Yukawa couplings
$$
y_{\a\b\g} = -\int_{\ca{W}} \O\wedge\partial_{\a\b\g}\O~,$$
where in this context $\a,\b,\g$ run over $\ps$ and $\ph$, can be computed
either from the
Picard--Fuchs equations or by means of a calculation in the ring of the
defining polynomial. The latter calculation proceeds by multiplying three
deformations of $p$ and reducing the result modulo the Jacobian ideal of
$p$.
The couplings are identified through the relation
$$
{\partial_\a p}\,{\partial_\b p}\,{\partial_\g p} \simeq y_{\a\b\g}
{{\bf h}\over \langle {\bf h} \rangle}$$
where ${\bf h}$ denotes the Hessian\Footnote{The Hessian of $p$ is
the determinant of the matrix of second derivatives of $p$.} of $p$ and
\hbox{$\langle {\bf h} \rangle$} is a purely numerical value (independent
of
the parameters) corresponding to ${\bf h}$ which may be computed via
considerations of topological field theory. This factor can also be fixed
from
our knowledge of the periods by means of the relation
$$
y_{\a\b\g} = - \ip^T\S\,\partial_{\a\b\g}\ip = -
\vp^T\s\,\partial_{\a\b\g}\vp
\eqlabel{yper}$$
with
$$
\S=\pmatrix{\- 0&\- {\bf 1}\cr
        -{\bf 1}&\- 0      \cr}
{}~~~~\hbox{and}~~~~\s=m^T\S m~.$$
The normalization is then also fixed by computing the couplings in the
limit of small $\ps$. Proceeding in either way we find
$$
\eqalign{y_{\ps\ps\ps}&= -{4096i\over \p^3}{\ps^5\over\D}~,\cropen{7pt}
         y_{\ps\ph\ph}&= -{  64i\over
\p^3}{\ps^3(\ph+4\ps^4)\over (1-\ph^2)\D}~,\cr}
\hskip30pt
\eqalign{y_{\ps\ps\ph}&= -{ 128i\over \p^3}{\ps^2\over \D}~,
\cropen{7pt}
         y_{\ph\ph\ph}&= -{   8i\over \p^3}
{\ps^4(1 + 3\ph^2 + 16\ps^4\ph)\over (1-\ph^2)^2\D}~,\cr}
\hskip30pt\boxed{$\ca{F}_0 = \vp_0$}$$
with again
$$
\D=(8\ps^4 + \ph)^2 - 1 ~.$$
The gauge in which this has been computed is the gauge with
\hbox{$\ca{F}_0=\vp_0$}. We compute now the couplings in the large complex
structure gauge in which \hbox{$\ca{F}_0=1$}. This introduces a factor of
\hbox{$1/\vp_0^2$} in addition to the usual tensor transformation rules.
Thus,
for example
$$
y_{ttt} ={1\over \vp_0^2}\left(
y_{\ps\ps\ps} \left({d\ps\over dt}\right)^3 +
3y_{\ps\ps\ph}\left({d\ps\over dt}\right)^2\left({d\ph\over dt}\right)+
3y_{\ps\ph\ph}\left({d\ps\over dt}\right)\left({d\ph\over dt}\right)^2+
y_{\ph\ph\ph}\left({d\ph\over dt}\right)^3\right)$$
On performing the expansion in powers of $q$ we find
$$\eqalign{
y_{ttt}&=
1 + 160 \cos (\pi  s) q^\half +
\Bigl( 72224 + 20224 \cos (2 \pi  s) \Bigr)  {q} +\cr
&\quad\Bigl( 50889600 \cos (\pi  s) +
     1946752 \cos (3 \pi  s) \Bigr)  {q^{3\over 2}} +\cr
&\quad\Bigl( 18774628384 +
   12120984320 \cos (2 \pi  s) + 175699968 \cos (4 \pi  s) \Bigr)  {q^2}
+\cr
&\quad\Bigl( 16276063840000 \cos (\pi  s) {+}
     2316635272000 \cos (3 \pi  s) {+}
     15477172160 \cos (5 \pi  s) \Bigr)  {q^{5\over 2}} {+}
\cdots\cropen{10pt}
y_{tts}&=
160 i \sin (\pi  s) q^\half + 20224 i \sin (2 \pi  s) {q} +
\Bigl( 16963200 i \sin (\pi  s) + 1946752 i \sin (3 \pi  s) \Bigr)
{q^{3\over 2}} +\cr
&\quad\Bigl( 6060492160 i \sin (2 \pi  s) +
     175699968 i \sin (4 \pi  s) \Bigr)  {q^2} +\cr
&\quad\Bigl( 3255212768000 i \sin (\pi  s){+}1389981163200 i \sin (3 \pi
s){+}
     15477172160 i \sin (5 \pi  s) \Bigr)  {q^{5\over 2}}{+}
\kern-1pt\cdots\cropen{10pt}
y_{tss}&=
-1 + 160 \cos (\pi  s) q^\half + 20224 \cos (2 \pi  s) {q} +
\Bigl( 5654400 \cos (\pi  s) {+} 1946752 \cos (3 \pi  s) \Bigr)
{q^{3\over 2}} +\cr
&\quad\Bigl( 3030246080 \cos (2 \pi  s) +
175699968 \cos (4 \pi  s) \Bigr)  {q^2} +\cr
&\quad\Bigl( 651042553600 \cos (\pi  s) + 833988697920 \cos (3 \pi  s) +
     15477172160 \cos (5 \pi  s) \Bigr)  {q^{5\over 2}} +
\cdots\cropen{10pt}
y_{sss}&=
2i\cot (\pi  s) + 160 i \sin (\pi  s) q^\half + 20224 i \sin (2 \pi  s)
q+\cr
&\quad\Bigl( 1884800 i \sin (\pi  s) +
1946752 i \sin (3 \pi  s) \Bigr)  {q^{3\over 2}} +\cr
&\quad\Bigl( 1515123040 i \sin (2 \pi  s) +
    175699968 i \sin (4 \pi  s) \Bigr)  {q^2} +\cr
&\quad\Bigl( 130208510720 i \sin (\pi  s) {+} 500393218752 i \sin (3 \pi
s){+}
     15477172160 i \sin (5 \pi  s) \Bigr)  {q^{5\over 2}} {+} \cdots\cr}
$$

We wish to interpret these results as an instanton sum. First however we
must
relate the classical part, $\yzero_{\a\b\g}$ of the couplings to the
intersection numbers of the homology basis $(H,L)$ of
Section~\chapref{manifold}.

The definition of the flat coordinates by means of \eqref{flatcds} amounts
to
the identification of the tangent space to the \K-cone with $H^2(\M)$,
explicitly we have
$$\eqalign{
\pd{}{t^1}&=H\cropen{10pt}
dt^1&={HL\over 4}=h\cr}\hskip30pt
\eqalign{
\pd{}{t^2}&=L\cropen{10pt}
dt^2&={H^2\over 4} - {HL\over 2}=l~.\cr}\eqlabel{tangsp}$$
In view of the linear relation \eqref{ts} between $(t^1,t^2)$ and the new
coordinates $(t,s)$ we have also
$$\eqalign{
\pd{}{t} &= {H\over 2}\cropen{10pt}
dt &={H^2\over 4}=2h + l\cr}\hskip30pt
\eqalign{
\pd{}{s} &= {H\over 2} - L\cropen{10pt}
ds &= - {H^2\over 4} + {HL\over 2}= - l~.\cr}$$
It is now simple to identify the classical part of the Yukawa couplings.
For
example we have
$$
\yzero_{ttt}~=~\yzero\left(\pd{}{t},\pd{}{t},\pd{}{t}\right)~=~
\left({H\over 2}\right)^3~.$$
Thus we find
$$
\yzero_{ttt}~=~1~~,~~\yzero_{tts}~=~0~~,~~\yzero_{tss}~=~-1~~,~~
\yzero_{sss}~=~-2$$
and we see that the first three values agree with the leading terms of the
Yukawa couplings computed above while the value for $\yzero_{sss}$ tells
us how
to divide the $\cot\,\p s$ term into a classical part and an instanton
part.
The fact that there is an instanton contribution to $\yzero_{sss}$ that
does
not go to zero in the limit $q\to 0$ is in part due to the fact that there
are
instantons that form a continuous family and also due to the fact that we
shall
have to revert to the coordinates $(t^1,t^2)$ in order to take the proper
limit
that suppresses the instanton contributions.

The interpretation of the couplings as instanton sums involves writing the
complex \K--form in terms of the coordinate basis
$$
B+iJ = t\pd{}{t} + s\pd{}{s}$$
and integrating over the holomorphic images of the worldsheet $\S$. These
images can be classified by homology type with respect to the basis
$(h,l)$:
$$\eqalign{
\S_{jk} &= jh + kl\cr
        &={j\over 2}dt + \left({j\over 2}-k\right)ds~.\cr}$$
The value of the action corresponding to such an instanton is
$$
S_{jk}= -2\p i\int_{\S_{jk}}(B+iJ) = -2\p i\left[ {j\over 2}t +
\left({j\over 2} - k\right)s\right]~.$$
Consistent with these observations we see that the series expansions for
the
couplings do indeed have the form
$$
y_{\a\b\g} = \yzero_{\a\b\g} +
\sum_{j,k}{c_{\a\b\g}(j,k) \, n_{jk}\,q^{j\over 2} r^{{j\over 2}-k}
\over 1-q^{j\over 2} r^{{j\over 2}-k}}~.\eqlabel{instantons} $$
Here we have written $r=e^{2\p i s}$ and the $c_{\a\b\g}$ are given by
$$
\pmatrix{c_{ttt}\cropen{3pt}
         c_{tts}\cropen{3pt}
         c_{tss}\cropen{3pt}
         c_{sss}\cropen{3pt}} =
{1\over 8}\pmatrix{j^3\cropen{3pt}
             j^2(j-2k)\cropen{3pt}
             j(j-2k)^2\cropen{3pt}
              (j-2k)^3\cropen{3pt}}~.$$
Note that the $2i\cot(\p s)$ term fits into this scheme since after
separating
out the classical piece we have
$$
2i\cot(\p s)  = - 2 - {4r^{-1}\over 1-r^{-1}}~.\eqlabel{cot}$$
Thus it would appear to correspond to a contribution of instantons of type
$(0,1)$.
With the exception of these we observe from our series expansions for the
couplings that the $n_{jk}$ vanish for $k>j$ and $n_{jk}=n_{j,j-k}$ for
$j\neq 0$. The values of the $n_{jk}$ for $j\leq 6$ are given in
\hbox{Table~\tabref{insts}}

\vbox{
$$\vbox{\offinterlineskip\halign{
\strut # height 15pt depth 8pt& \quad$#$\quad\vrule
&\quad$#$\quad\hfil\vrule&\quad$#$\quad\hfil\vrule&\quad$#$\quad\hfil\vrule

&\quad$#$\quad\hfil\vrule
\cr
\noalign{\hrule}
\vrule&j\hfil&\hfil k=0&\hfil k=1&\hfil k=2&\hfil k=3\cr
\noalign{\hrule}
\vrule&0&0           & 4             & 0              & 0 \cr
\noalign{\hrule}
\vrule&1&640         & 640           & 0              & 0 \cr
\vrule&2&10032       & 72224         & 10032          & 0 \cr
\vrule&3&288384      & 7539200       & 7539200        & 288384  \cr
\vrule&4&10979984    & 757561520     & 2346819520     & 757561520  \cr
\vrule&5&495269504   & 74132328704   & 520834042880   & 520834042880  \cr
\vrule&6&24945542832 & 7117563990784 & 95728361673920 & 212132862927264 \cr
\noalign{\hrule}
}}
$$
\nobreak\tablecaption{insts}{A table of numbers, $n_{jk}$, of
instantons of type \hbox{$(j,k)$} for \Mone.
The numbers $n_{0k}$ are
special, $n_{01}$ being the only nonzero number of this type.
For $j\geq 1$ and $0\leq k\leq \left[ j/2 \right]$ we have
\hbox{$n_{jk}=n_{j,j-k}$}. For $k>j$ the $n_{jk}$ vanish. }}
\vskip15pt
\noindent We may understand the fact that for $j\ge 1$ the instanton
numbers satisfy $n_{jk}=n_{j,j-k}$ as a consequence of a monodromy
transformation $\ca{T}_\infty$ on the coordinates $(t^1,t^2)$. It is easy
to check that $\ca{T}_\infty$ has the effect
$$
\ca{T}_\infty:~(t^1,t^2)\to (t^1 - t^2,\, - t^2 + 1)~.$$
It follows that $(dt^1,dt^2)\to (dt^1 - dt^2, -dt^2)$ and in virtue of the
identification \eqref{tangsp} that $(h,l)\to (h+l, -l)$ which in turn
has the consequence $(j,k)\to (j,j-k)$. In other words it is being asserted
that consistency with mirror symmetry requires the existence of a monodromy
transformation on the \K--class
parameters of \ca{M} the effect of which is $(h,l)\to (h+l, -l)$. It is
interesting to note how $n_{01}$ term escapes this argument. The effect of
$\ca{T}_\infty$ on $s$ is $s\to -s-1$ so the effect on the coupling
$y_{sss}$
is $y_{sss}\to -y_{sss}$. Note that $\cot\,\p s$ transforms appropriately
but
if we separate this term into a classical part and an instanton part as in
\eqref{cot} then the classical part changes sign in virtue of the
transformation
$$
-{4r^{-1}\over 1 - r^{-1}}\to +{4r^{-1}\over 1 - r^{-1}} + 4$$
the effect being that the transformation of the instanton part changes the
classical part.

The appearance of the instanton sum can be improved, and the nature of the
large complex structure limit clarified, by reverting to the $(t^1,t^2)$
coordinates.
Now we have $\S_{jk}=jdt^1 + kdt^2$. We write also
$$
q_1\define e^{2\p i t^1}= \sqrt{q\over r}~~~,
{}~~~q_2\define e^{2\p i t^2}={1\over r}~.$$
The instanton sum contains now only positive powers of $q_1$ and $q_2$
$$
y_{\a\b\g} = \yzero_{\a\b\g} +
\sum_{j,k}{c_{\a\b\g}(j,k) \, n_{jk}\,q_1^j q_2^k
\over 1 - q_1^j q_2^k}~,~~~~~~
\pmatrix{\yzero_{111}\cr \yzero_{112}\cr \yzero_{122}\cr \yzero_{222}\cr}
= \pmatrix{8\cr 4\cr 0\cr 0\cr}~.
\eqlabel{instantons2} $$
With respect to the new basis the quantities $c_{\a\b\g}$ also simplify
$$
\pmatrix{c_{111}\cr
         c_{112}\cr
         c_{122}\cr
         c_{222}\cr} =
\pmatrix{j^3\cr
         j^2 k\cr
         j   k^2\cr
             k^3\cr}~. \eqlabel{cijk}$$
The large complex structure limit is evidently the limit
\hbox{$(q_1,q_2)\to (0,0)$}.

Let us now consider \Mtwo . The Yukawa couplings can be obtained
from the ring of the defining polynomial and their normalizations
fixed by the asymptotic behavior of the periods as we have explained.
We find in this case
$$
\eqalign{
y_{\ps\ps\ps}&= -{i\over 64\p^3}{(12\ps)^9\over\D}~,\cropen{7pt}
y_{\ps\ps\ph}&= -{  3i\over 4\p^3}{(12\ps)^4\over \D}~,\cr}
\hskip30pt
\eqalign{y_{\ps\ph\ph}&=
-{ i\over 48\p^3}{(\ph + 432 \ps^6)(12\ps)^5\over (1 - \ph^2)\D}~,
\cropen{7pt}
         y_{\ph\ph\ph}&= -{ i\over 6912\p^3}
{(1 + 3\ph^2 + 1728\ps^6\ph)(12\ps)^6\over (1-\ph^2)^2\D}~,\cr} $$
with
$$
\D=(864\,\ps^6 + \ph)^2 - 1 ~.$$
These couplings have been computed in the ${\cal F}_0 = \vp_0$ gauge.

{}From the inverse mirror map we can compute the $y_{\a\b\g}$ in the
large radius gauge ${\cal F}_0=1$. The resulting instanton
sum is of the expected form \eqref{instantons2} with the $c_{\a\b\g}$
given in \eqref{cijk}. Values for the instanton numbers are
displayed in \hbox{Table~\tabref{inst12}}.
As in \Mone , the numbers $n_{0k}$ are
special, $n_{01}$ being the only nonzero number of this type.
For $j\geq 1$ we again have
the quantum symmetry \hbox{$n_{jk}=n_{j,j-k}$}.
Also, the $n_{jk}$ vanish for $k>j$.

\vskip25pt
\vbox{
$$\vbox{\offinterlineskip\halign{
\strut # height 15pt depth 8pt& \quad$#$\quad\vrule
&\quad$#$\quad\hfil\vrule&\quad$#$\quad\hfil\vrule&\quad$#$\quad\hfil\vrule
\cr
\noalign{\hrule}
\vrule&j\hfil&\hfil k=0&\hfil k=1&\hfil k=2           \cr
\noalign{\hrule}
\vrule&0&0           & 2             & 0              \cr
\noalign{\hrule}
\vrule&1&2496        & 2496          & 0              \cr
\vrule&2&223752      & 1941264       & 223752         \cr
\vrule&3&38637504    & 1327392512    & 1327392512     \cr
\vrule&4&9100224984  & 861202986072  & 2859010142112  \cr
\noalign{\hrule}
}}
$$
\nobreak\tablecaption{inst12}{A table of numbers, $n_{jk}$, of
instantons of type \hbox{$(j,k)$} for \Mtwo. }}
\vskip30pt
\subsection{Instantons of genus one}
Following Bershadsky {\it et al.\/}
\cite{\rBCOV}\
we consider the quantity $F_1$
defined by a certain path integral and whose topological limit
is given by an expression of the form
$$ F_1^{top} = \log \left[ \left({\ps\,d\over
\vp_0}\right)^{3+b_{11}-\chi/12}
\, {\partial(\ps,\ph)\over \partial(t^1,t^2)}\,f\right] +
\hbox{const.} ~, \eqlabel{topF}$$
We recognise $1/\vp_0$ as a gauge factor. The additional factor of
$\ps^{3+b_{11}-\chi/12}$ is needed to compensate for the term
$e^K$ in eq.~(16) of \cite{\rBCOV}, which grows like $|\ps|^{-2}$ near
$\ps=0$.  Here $K$ as usual is the K\"ahler potential.  The asymptotic
growth of $e^K$ follows immediately from the factor of $\ps$ in the expression
for $\O$, since the locus $\ps=0$ parametrizes nonsingular manifolds.
When $b_{11}=2$, the exponent becomes \hbox{$5-\ch/12$}.
Then there is a Jacobian and a holomorphic function $f$
which is determined by the conditions that $F_1^{top}$ can be
singular
only when $\D=0$ or $\ph^2=1$ together with the condition on the large
radius
limit. The relevance of $F_1^{top}$  to us here is that in virtue
of
mirror symmetry it enjoys an expansion
$$ F_1^{top} = -{2\p i\over 12}c_2\cdot(B+iJ) + \hbox{const.} -
\,\sum_{jk}\left[ 2 d_{jk}\log\eta(q_1^j q_2^k) +
{1\over 6}n_{jk}\log(1 - q_1^j q_2^k) \right]~, \eqlabel{Finst}$$
where
$$
\eta(q) = \prod_{n=1}^\infty (1-q^n)$$
is the Dedekind function.
In the expansion for $F^{top}$ the leading
term is product of the second Chern class with the \K-class and determines
the asymptotic form of $F_1^{top}$ in the large radius limit. The summation in
\eqref{Finst} is over instanton contributions and $d_{jk}$ and $n_{jk}$
are the numbers of instantons of genus one and genus zero.

The quantity $\ps/\vp_0$ is regular and nonzero
at $\ps=0$ so the function $f$ must be of the form
$$
f=\D^a\,(\ph^2 - 1)^b\, \ps^c\eqlabel{fhol}$$
The exponent $c$ is fixed by the behavior of the Jacobian
${\partial(\ps,\ph)\over \partial(t^1,t^2)}$
at $\ps=0$. The remaining exponents are then
determined by comparing \eqref{topF} with the
leading term in \eqref{Finst} in the large radius limit
$$\eqalign{
F_1^{top}&\asymp -{2\p i\over 12}c_2\cdot(B+iJ)\cropen{10pt}
         &= -{2\p i\over 12}c_2\cdot(t^1H +t^2L)  \cropen{10pt}
         &= -{2\pi i\over 12}
\left\{ \eqalign{56t^1 + 24t^2~~&\hbox{for}~~\Mone\cr
                 52t^1 + 24t^2~~&\hbox{for}~~\Mtwo\cr}\right.
\cr} \eqlabel{limF}
$$
On the other hand, from \eqref{topF}, \eqref{fhol} and the form
of the mirror map, we see that
$$
F_1^{top} \asymp -{2\pi i\over 12}\left\{
\left[{2\over d}(72 - \chi+ 12c) + 24a\right]\,  t^1
+ \left[6 + 12a + 12b + {1\over d}(72 - \chi + 12c)\right]\,
 t^2\right\} ~,$$
allowing us to fix the constants $a$ and $b$ once $c$ is known.

For \Mone\ we have, $c$=0, $a=-1/6$ and $b=-5/6$ while
for \Mtwo\ we have, $c=1$, $a=-1/6$ and $b=-2/3$.
Substituting the corresponding holomorphic function \eqref{fhol}
in \eqref{topF} and using our previous results
we find the instanton numbers. In both models they satisfy the
quantum symmetry \hbox{$d_{jk}=d_{j,j-k}$}. Also, $d_{0k}=0$ and
\hbox{$d_{jk}=0$ for $k > j$. }
Tables of values are given below.
\newpage
\vskip15pt
\vbox{
$$\vbox{\offinterlineskip\halign{
\strut # height 15pt depth 8pt& \quad$#$\quad\vrule
&\quad$#$\quad\hfil\vrule&\quad$#$\quad\hfil\vrule&\quad$#$\quad\hfil\vrule
&\quad$#$\quad\hfil\vrule
\cr
\noalign{\hrule}
\vrule&j\hfil&\hfil k=0&\hfil k=1&\hfil k=2&\hfil k=3\cr
\noalign{\hrule}
\vrule &1&\- 0&0&0&\- 0\cr
\vrule &2&\- 0&0&0&\- 0\cr
\vrule &3& -1280&2560&2560&-1280\cr
\vrule &4& -317864&1047280&15948240&\- 1047280\cr
\vrule &5& -36571904&224877056&12229001216&\- 12229001216\cr
\vrule &6& -3478899872&36389051520&4954131766464&\- 13714937870784\cr
\noalign{\hrule}
}}
$$
\nobreak\tablecaption{dinsts}{A table of numbers, $d_{jk}$, of genus one
instantons of type \hbox{$(j,k)$} for \Mone. The numbers displayed are for
$j=1,\ldots,6$ and $k\le \left[j/2\right]$. We have
\hbox{$d_{jk}=d_{j,j-k}$} and \hbox{$d_{jk}=0$} for $k > j$.} }
\vskip35pt
\vbox{
$$\vbox{\offinterlineskip\halign{
\strut # height 15pt depth 8pt& \quad$#$\quad\vrule
&\quad$#$\quad\hfil\vrule&\quad$#$\quad\hfil\vrule&\quad$#$\quad\hfil\vrule
\cr
\noalign{\hrule}
\vrule&j\hfil&\hfil k=0&\hfil k=1&\hfil k=2           \cr
\noalign{\hrule}
\vrule&1&0           & 0             & 0              \cr
\vrule&2&-492        & 480           & -492           \cr
\vrule&3&-1465984    & 2080000       & 2080000        \cr
\vrule&4&-1042943028 & 3453856440    & 74453838480    \cr
\noalign{\hrule}
}}
$$
\nobreak\tablecaption{dinst12}{A table of numbers, $d_{jk}$, of
genus one
instantons of type \hbox{$(j,k)$} for \Mtwo. }}
\newpage
\section{verification}{Verification of Some Instanton Contributions}
In this section, we verify the predictions for some of the numbers of
rational and elliptic curves given by our calculation of the Yukawa couplings.
In the process, we
consider more generally a \cy\ manifold $\ca{M}$ which contains
an exceptional divisor $E$ which is a $\IP^1$ bundle
over a smooth curve $C$ of genus at least 2.
More precisely, we suppose that
there is a canonical threefold $\Mhat$ and a birational map $f:\ca{M}\to\Mhat$
such
that $f(E)=C$, with $f$ an isomorphism outside $E$ and $C$.  It follows
that the fibers of $f$ over points of $C$ are smooth rational curves $l$
which satisfy $l\cdot E=-2$.

For any curve $\G\subset\ca{M}$, we write $n_\G=\G\cdot E$.
Suppose in addition that $\G$ is an irreducible rational curve.
If $\G\subset E$,
then $\G$ must be one of the fibers $l$.  If $\G\not\subset E$,
note that $n=n_\G\ge 0$.
For each of the $n$ intersection points $p_i$ of $\G$ and $E$,
there is a unique fiber $l_i$ of $E$ which meets that point.  Taking the union
of $\G$ with all of these, we arrive at an {\sl connected} curve
$\G'$ which is homologous to $\G+nl$.  This has an interpretation as a
degenerate
instanton, and as such
contributes to the instanton sum (see appendix to \cite{\rBCOV}).
In the sequel, a {\sl degenerate instanton} denotes a connected
curve of arithmetic genus 0.

This can be generalized to all degenerate instantons as follows.  First of
all, we assume that we are in a generic situation so that
\bigskip
\item{1.~} There are only
finitely many irreducible rational curves in a given homology class (with the
exception of the homology class of $l$).
\item{2.~} The images of any two irreducible rational curves via $f$, neither
equal to a fiber of $f$, are disjoint in $\Mhat$.
\bigskip
\noindent Let $\G$ now denote any connected degenerate
instanton.  Suppose $\G$ contains $m$ fibers $l_1,\ldots l_m$
of $f$ as components.  Then
$\tilde{\G}=\G-(l_1+\ldots +l_m)$ is connected.  Put $n=n_{\tilde{\G}}\ge 0$.
The $m$ fibers $l_i$ meet $\tilde{\G}$ in $m$ of the $n$ points
$\tilde{\G}\cap E$.  To the remaining $n-m$ points are associated
fibers $l_{m+1},\ldots l_n$.  Put $\G'=\tilde{\G}+l_{m+1}+\ldots +l_n$.
The association $\G\mapsto \G'$ clearly gives an involution on the set
of all (possibly degenerate) instantons, with the exception of the fibers.
Note that whether or not $\G$ is irreducible, we have
that $\G'$ is homologous to $\G+(\G\cdot E)l$.

Let $T_{\ca{M}}$ denote the holomorphic tangent bundle of $\ca{M}$.
The moduli space of complex structures of $\ca{M}$ has dimension
$h^{2,1}(\ca{M})$,
and its tangent space is canonically isomorphic to $H^1(T_{\ca{M}})$.
$H^{1,0}(C)$ contributes a summand to $H^1(T_{\ca{M}})$
(in the cases of $\wP[8]$ and $\WP[12]$,
this complements the space of polynomial deformations
of $\ca{M}$).  For each
fiber $l$, restriction defines a map $r:H^1(T_{\ca{M}})\to
H^1(N_{l/{\ca{M}}})$,
where
$N_{l/{\ca{M}}}$ denotes the normal bundle of $l$ in \ca{M}\
(which is $\ca{O}\oplus\ca{O}(-2)$ by our assumptions).
Deformation
theory tells us that in order for one of the lines $l$ to be  holomorphically
deformable to first order when the complex structure is perturbed in the
direction of an
element $\rho\in H^1(T_{\ca{M}})$, it is necessary
and sufficient that $r(\rho)=0$.
In addition, we calculate that $H^1(N_{l/{\ca{M}}})\simeq\IC $ globalizes
to the canonical bundle $K_C$ of $C$.  There results a map
$H^1(T_{\ca{M}})\to H^0(K_C)$, also denoted $r$.  Said differently, to each
first order deformation $\rho\in H^1(T_{\ca{M}})$ one gets a section $s$ of
$K_C$.
The lines that deform in the
direction of $\rho$ to first order are precisely those
lying over the points of $C$ at which $s$ vanishes.  In fact, we can show
more: if any 1~parameter deformation of \ca{M}\ is given, and
$\rho\in H^1(T_{\ca{M}})$ corresponds to the first order part of this
deformation,
then if $r(\rho)\in H^0(K_C)$ has distinct zeros, it follows that the
fibers over the points of $C$ at which $r(\rho)=0$ actually deform
holomorphically with our 1~parameter family.
Thus
$2g-2$ fibers deform, where $g$ is the genus of $C$.  The first order
calculation was originally done by P.M.H.~Wilson~\cite{\rWilson},
and was supplemented by a proof that at least one of the fibers deform.
The same result that
$2g-2$ is the instanton contribution for this family of curves could have
been obtained by the topological field theory argument of \
\REFS{\rAM}{P.~S.~Aspinwall and D.~R.~Morrison, Comm.\ Math.\ Phys.\
{\bf 151} (1993) 245.}
\refsend.

Now, deform ${\ca{M}}$ so that $E$ disappears, leaving only $2g-2$ fibers in
its
place.  These curves all have normal bundle $\ca{O}(-1)\oplus\ca{O}(-1)$, so
may
be simultaneously flopped, yielding a topologically distinct threefold
${\ca{M}}'$.  These curves will still be referred to as fibers in the sequel.

We now look at the effect of this flop on the Yukawa couplings.  For any
divisor class $H$ on ${\ca{M}}$, let $H'=H+(H\cdot l)E$.  The map $H\mapsto H'$
is the adjoint of the previously discussed involution
$\G\mapsto \G+(\G\cdot E)l$ on homology classes of
curves, that is $H\cdot \G=H'\cdot \G'$ for all $H,\G$.  This involution
was also considered in \cite{\rWilson}.

Let $H_1,H_2,H_3$ be any three $(1,1)$ forms.  We claim that the Yukawa
couplings satisy
$$
<H_1,H_2,H_3>_{\ca{M}}=<H_1',H_2',H_3'>_{{\ca{M}}'} \eqlabel{3pt}
$$

There are two types of terms in the expansion of the Yukawa couplings%
---the classical
term and the instanton terms.  The instanton terms match up for all instantons
except the fibers, by the above discussion.  And the combination of the
classical term with the contribution of the fibers to the instanton sum
can also be seen to match up by the form of the instanton term and the
relationship between intersection numbers and flops.  This calculation
has been done in a more general setting in~
\REF{\rphases}{E. Witten, ``Phases of $N=2$ Theories in Two Dimensions'',
preprint IASSNS-HEP-93/3.}
\cite{{\rAGM,\rphases}}.

Back to the cases $\wP[8]$ and $\WP[12]$,
the above discussion yields three conclusions, all of
which are borne out by the calculation of the Yukawa couplings.

\item{1.~} $n_{01}=4$ for $\wP[8]$ and $n_{01}=2$ for $\WP[12]$.
\item{2.~} $n_{jk}=n_{j,j-k}$ for $j\neq 0$.
\item{3.~} $n_{jk}=0$ for $j>0$ and $k>j$.
\item{4.~} The Yukawa couplings are the same for ${\ca{M}}$ and ${\ca{M}}'$.

To see (3), we let $\G$ be any
degenerate instanton with $\G\cdot H=j$ and $\G\cdot L=k$.  We write
$\G=\G_0\cup_{i=1}^nl_i$, where the $l_i$ are fibers and $\G_0$ is irreducible,
as discussed earlier.  Put $\G_0\cdot L=r$.  Then since $\G_0\not\subset E$,
it follows that
$$\G_0\cdot E=\G_0\cdot(H-2L)=j-2r\ge 0.$$
We also have
$n\le \G_0\cdot E=j-2r$.  Hence
$$k=\G\cdot L=\G_0\cdot L+n=r+n\le j-r\le j.$$

Finally, some of the numbers $n_{jk}$ for $j>0$ can be verified.  We begin
with the consideration of $\wP[8]$.
We let
the volume of the fibers shrink to 0 by letting $t^2\to 0$ inside the
K\"ahler moduli space.  This is accomplished by letting $q_2=1$.
The resulting 1~parameter coupling is seen to be the same as the coupling
for a $(2,4)$ complete intersection in $\IP^5$.  This is because there
is a natural mapping $\wP \to\IP^5$ given by
$$
(x_1,\ldots,x_5)\mapsto(y_0,\ldots,y_5)=(x_1^2,x_1x_2,x_2^2,x_3,x_4,x_5)
$$
The image $\Mhat$ clearly satisfies the quadratic equation $y_0y_2=y_1^2$, and
the weighted octic equation of $X$ becomes a quartic equation in the
$y_i$.  This singular threefold has one K\"ahler modulus; and since it
is a complex deformation of a smooth $(2,4)$ complete intersection, the
Yukawa coupling calculated here must coincide with that of the $(2,4)$
complete intersection.  In fact, we can see directly that the restricted
Picard-Fuchs equation or power series expansion of $\vp_0$ coincides with
that first found for this complete intersection by Libgober and Teitelbaum
\cite{\rLT}~
(see also \ \cite{{\rBvS,\rperiods}}).
Note also that the resulting relationship between the two parameter and one
parameter conformal field theories has already been noted---by means of the
mirror families---in Section 3.2.

To tie together the instanton numbers, we can establish the validity
of the following assertion.
Let $\Mhat$ be the singular model of ${\ca{M}}$, and let $\G\subset\Mhat$
be a rational
curve.  Let $\tilde{\G}\subset {\ca{M}}$ be the proper transform of $\G$.  Then
as $\Mhat$ deforms to a smooth $(2,4)$ complete intersection, $\G$ splits up
into $2^n$ distinct curves, where $n=n_{\tilde{\G}}$.  This is proven
by a local calculation and a local to global argument.  Note that this
result is compatible with the description of degenerate instantons, as there
are $2^n$ subsets of the $n$ fibers that meet $\tilde{\G}$ which can be joined
to $\tilde{\G}$ to yield a degenerate instanton.  A
consequence is that $\sum_kn_{jk}$ is the number of degree $j$
rational curves on
a $(2,4)$ complete intersection.  Combining
with the symmetry $n_{jk}=n_{j,j-k}$, we get the correct numbers for
$n_{10}=n_{11}$, since the number of lines on a $(2,4)$ complete intersection
in $\IP^5$ is well known to be 1280.
Alternatively, $n_{10}$ can also be verified by counting lines
on a pencil of K3 surfaces.  The linear system $|L|$ defines a fibration
of \ca{M}\ by K3 surfaces, which can be identified as quartic
K3 surfaces in $\IP^3$, as we have seen earlier.  So this pencil of
K3's can be related to a family of quartic surfaces, each of
which spans a $\IP^3$.
This can be accomplished by using the model contained in the blowup of
$\IP^5$ along the $\IP^2$ defined by $y_0=y_1=y_2=0$.  We use the linear system
$|2L|$ to get a map to the conic $y_0y_2=y_1^2$, and count lines in the
fibers.  We will get the same number by degenerating the conic to a union
of two lines; in other words, we replace \ca{M}\ be the complete intersection
of a quartic with a union of two hyperplanes formed out of $y_0,y_1,y_2$.
To count lines in these fibers, we see that the result is the same as
counting lines in $\IP^4$ which lie in a quartic hypersurface and {\sl some}
member of a linear system of hyperplanes; then multiplying this result by 2.
This can be calculated by standard techniques in enumerative geometry
(it is $2(c_5(S^4Q))(c_1(Q))$, where $Q$ is the universal quotient bundle on
the Grassmannian $G(2,5)$ of lines in $\IP^4$), and
we verify $n_{10}=640$.
The $n_{2k}$ numbers can also be calculated
by similar considerations, but now it is essential to calculate $n_{20}$
directly by the K3 pencil method.   By similar reasoning, $n_{20}$ is
twice the number of conics that are contained in a quartic hypersurface
in $\IP^4$ and some hyperplane of a varying pencil.  This is straightforward
to compute, and the result is that we verify $n_{20}=10032$.
This also verifies $n_{02}$, and $n_{11}$ is verified by comparing
$n_{20}+n_{11}+n_{02}$ with
the total number of conics on a $(2,4)$ complete intersection
in $\IP^5$, verified by \
\REFS{\rSvS}{S.~A.~Str\char"1C mme and D.~van~Straten, Private communication.}
\refsend.
The number of twisted cubics on a
$(2,4)$ complete intersection has been calculated by Ellingsrud and
Str\char"1C mme \
\Ref{\rES}{G.~Ellingsrud and S.~A.~Str\char"1C mme, In preparation.},
and agrees with $\sum_{k=0}^3n_{3k}$.  This number had
previously been predicted in \cite{{\rLT,\rBvS}}.

The situation for $\WP[12]$ is entirely analogous.
For instance, for $n_{10}$,
we take twice the number of lines in $\IP_5^{(1,1,1,1,1,3)}$
which lie on a fixed degree 6
hypersurface and some member of a pencil of linear hypersurfaces.
The calculation may be done by a modification of the technique in~
\Ref{\rKatz}{S.~Katz, ``Rational Curves on Calabi-Yau Manifolds: Verifying
Predictions of Mirror Symmetry'', OSU-M-92-3.}.
The result is $n_{10}=2496$. The numbers of rational curves of low degree for a
$(2,6)$ intersection in $\IP_5^{(1,1,1,1,1,3)}$ have also been calculated
recently in Ref.~
\Ref{\rKTbis}{A.~Klemm and S.~Theisen, ``Mirror Maps and Instanton Sums for
Complete Intersections in Weighted Projective Space'', LMU-TPW-93-08 (1993).}.
These numbers agree with our results for $\sum_{k=0}^j n_{jk}$, as well as
with our direct calculation of these numbers by standard techniques in
enumerative geometry.

Turning to the verification of the predicted numbers of elliptic curves in
$\wP[8]$,
we use the K3 pencil method.  There are no projective elliptic curves of degree
less than 3, verifying the first two lines of \hbox{Table~\tabref{dinsts}}.
To verify $d_{30}$, we ``count'' plane cubic curves in $\IP^4$ which lie in a
quartic hypersurface and some member of a linear system of hyperplanes, then
multiply the resulting number by 2.  Here there is actually an infinite
number: the supporting plane of the cubic curve meets the quartic residually
in a line, and conversely, given any line on the quartic also in a hyperplane
of a pencil, the pencil of hyperplanes in the $\IP^3$ through that line
residually defines a 1~parameter family of cubics.  Using a formal
application of standard techniques
of intersection theory in enumerative geometry (on the moduli space of plane
cubic curves, multiply $c_{12}$ of the bundle of quartic forms restricted to
the varying curves by $c_3$ of the bundle of linear forms), the resulting
number is
indeed $-1280$.  Alternatively, this number could have been calculated by
a topological field theoretic argument; since each of the 640 lines found
in the $n_{30}$ calculation corresponds to a family of cubic curves
parametrized by $\IP^1$, the upshot is that each contributes
\hbox{$c_1(\IP^1)=-2$},
again arriving at $d_{30}=-1280$.  $d_{31}$ is verified as in the verification
of $n_{11}$, by relating $d_{30} + \ldots + d_{03}$ to the number of elliptic
cubics on a $(2,4)$ complete intersection in $\IP^5$, which can be easily
calculated to be 2560.

Considerations similar to the foregoing apply to \Mtwo.  In particular,
to see that $d_{10}=d_{11}=0$, we look at the blown down model obtained
by applying
the morphism determined by $|H|$. This maps $\ca{M}$ onto a quadric in
$\IP^4$, and
exhibits $\Mhat$ as a double cover of the quadric branched along a sextic
(the degree 12 equation for the weighted hypersurface becomes sextic in the
variables of $\IP^4$, and the additional $x_5^2$ term gives rise to the
double cover).
A curve $C$ satisfying $C\cdot H=1$ necessarily maps to a line in $\IP^4$
and must map birationally onto the line.  Such a curve would then have genus
0, and in particular not be elliptic.

We have observed that a negative instanton contribution can only arise when
there are continuous families of instantons for generic parameter values.
Continuous families can sometimes be avoided by doing the
calculations on nearby almost complex manifolds instead.  In fact, for
rational curves,
McDuff's transversality theorem~
\Ref{\rMcDuff}{D. McDuff, Invent. Math. {\bf 89} (1987) 13.}\
shows that continuous families cannot exist for the generic almost complex
structure; this is why all of the $n_{jk}$ are positive (even when there were
continuous families of instantons for general values of the parameters).  It is
therefore all the more striking that negative instanton values {\sl do}
occur for elliptic curves.
\vskip1in
\acknowledgements
It is a pleasure to thank Per Berglund, Sergio Ferrara, Wolfgang Lerche, Jan
Louis and Fernando Quevedo for instructive discussions. A.~F. thanks the Centro
Cient\'\i fico IBM-Venezuela for the use of its facilities and the Institut de
Physique of the Universit\'e\ de Neuch\^atel as well as ICTP-Trieste for
hospitality while part of this research was done.
\newpage
\immediate\closeout\referencewrite\referenceopenfalse
\line{\bf\hfil References\hfil}\bigskip\parindent=0pt\input referenc.texauxil
\bye